\documentclass[a4paper, 10pt, oneside]{article}
\usepackage[latin1]{inputenc}
\usepackage{amsmath}
\usepackage{amsfonts}
\usepackage{amssymb}
\usepackage{graphicx}
\usepackage{bigints}
\usepackage{verbatim}
\usepackage{url}
\usepackage{bm}
\usepackage{bbm}
\usepackage{float}
\usepackage{pdflscape}
\DeclareMathAlphabet{\mathpzc}{OT1}{pzc}{m}{it}
\newcommand{\dd}{\: \mathrm{d}}
\renewcommand{\P}{\mathbb{P}}
\newcommand{\E}{\mathbb{E}}
\newcommand{\Keywords}[1]{\par {\small{\bf Keywords\/}: #1}}

\title{Modeling catastrophic deaths using EVT with a microsimulation approach to reinsurance pricing}
\author{Matias Leppisaari\footnote{Model IT Ltd, Unioninkatu 13, Helsinki, Finland and Department of Mathematics and Systems Analysis, Aalto University. E-mail: matias.leppisaari@modelit.fi}}
\date{This version: October 30, 2013}

\begin{document}
\maketitle

\begin{abstract}
Recently, a marked Poisson process (MPP) model for life catastrophe risk was proposed in \cite{EKH:12}. We provide a justification and further support for the model by considering more general Poisson point processes in the context of extreme value theory (EVT), and basing the choice of model on statistical tests and model comparisons. A case study examining accidental deaths in the Finnish population is provided.

We further extend the applicability of the catastrophe risk model by considering small and big accidents separately; the resulting combined MPP model can flexibly capture the whole range of accidental death counts. Using the proposed model, we present a simulation framework for pricing (life) catastrophe reinsurance, based on modeling the underlying policies at individual contract level. The accidents are first simulated at population level, and their effect on a specific insurance company is then determined by explicitly simulating the resulting insured deaths. The proposed microsimulation approach can potentially lead to more accurate results than the traditional methods, and to a better view of risk, as it can make use of all the information available to the re/insurer and can explicitly accommodate even complex re/insurance terms and product features. As an example we price several excess reinsurance contracts. The proposed simulation model is also suitable for solvency assessment.
\end{abstract}


\section{Introduction} \label{introduction}
The assessment of catastrophic death risk is important for life insurers, and for non-life and health insurers writing accidental death covers. A good view of this risk is needed to explicitly assess the need for reinsurance, and to establish the amount of capital required to cover the risks left at the company's own account. Conversely, the reinsurer needs to be able to assess the catastrophe risk quantitatively in order to set an adequate -- but competitive -- price for the reinsurance cover. A model for catastrophe risk is also needed in the design of a reinsurance programme, to assess the effect of different proposed features, such as aggregate limits or drop-down features.

The measurement of catastrophe risks as well as the suitability and likely outcomes of different plans of action -- such as acquiring reinsurance, reserving capital, or using insurance linked securities to trasfer risk to capital markets -- require appropriate quantitative tools. In this paper we propose one tool for addressing the catastrophe risk in life and (health/non-life) accident insurance. Our approach is based on a population level model for occurrence of catastrophe events, combined with policy level simulation of the effects the catastrophes have on a specific insurance portfolio: The times and sizes of accidents are modeled using Poisson point process specification, followed by first modeling the propotion of total lives lost that had a (certain type of) insurance policy (based on national insurance coverage statistics), and finally the proportion of insured lives lost that were customers of a specific insurance company (based on market share statistics). Based on the accidental death model, we propose a novel application of a policy-by-policy simulation framework to life catastrophe risk measurement and reinsurance pricing; a concrete implementation of the simulation framework is made possible by the catastrophe model.

Catastrophe risk models are now commonly used in property insurance, where there are widely acknowledged commercial vendor models in use (e.g., AIR, RMS, EQECAT).\footnote{Publicly available information on the proprietary models can be found at \url{www.air-worldwide.com}, \url{www.rms.com}, and \url{www.eqecat.com}.} These models generally simulate occurrence times, intensity, geographical location, and possible other relevant characteristics of events causing catastrophes, and then estimate the effect these events have on the insurance company through the policies it has written. The approach works particularly well when we examine losses caused by a specific, well-defined phenomenon. Examples of these are natural phenomena, such as earthquakes, hurricanes, windstorms and floods. These events are caused by underlying physical processes, and can often be modeled with good or moderate success based on the scientific knowledge about the characteristics of the processes generating them. Indeed, the commercial catastrophe risk models are mainly intended for modeling losses caused by natural disasters.

With life policies the situation is more complicated in that the death covers written by life insurers generally cover all causes of death. This means that there is no well-defined process responsible for causing the insured losses of a company, and hence the setting does not lend itself to the kind of modeling adopted with natural perils. An alternative is to assess the number of deaths using purely statistical methods, without explicitly modeling the causes of deaths. This is the approach taken in this paper.

The statistical approach to catastrophes has of course its own problems, the biggest being the scarcity of relevant data about extreme events. Furthermore, our ultimate aim is not only to describe the observed accidental deaths by means of a statistical model, but to use the model for extrapolating outside the range of data: i.e., to assess the probabilities of events more extreme than those observed so far. This requires well-founded probabilistic and statistical methods, backed by a sound mathematical theory: we use extreme value theory (EVT) (\cite{EKM:97, COL:01, LLR:83}) -- in particular, the point process view of extremes (\cite{RES:87, SMI:89}) -- in order to develop a plausible population level model for accidental death events.

To go from population level risk to insurance company specific risk, we leverage the special structure of the insurance products considered. With products covering the death of an insured, the insured event is binary: the customer either dies or does not die -- there are no partial losses, in contrary to non-life insurance. In addition, the risk sum payable given the time of death is usually either known or directly calculatable based on contract terms (in risk and non-participating traditional savings policies); or it can be reasonably estimated e.g. as part of a policy level simulation model (for with-profit and unit-linked savings policies). It is our view that these features of life and personal accident products call for an explicit simulation of catastrophic events (numbers of deaths), combined with simulating the effect the event has on the insurance company under consideration, given the data regarding all its policies in force.

The standard intensity-based approaches to modeling mortality in life insurance are not suitable for modeling catastrophes, and hence we need to use techniques more in line with those applied in non-life insurance (cf. \cite{DPP:94}). Although the use of traditional reinsurance pricing methods -- experience rating and exposure rating, see \cite{GWP:06} and the references therein -- is still prevalent in practice, and these methods have their place, it has long been recognized that (proper) pricing of some features of reinsurance contracts requires a simulation approach. Unlike in (non-life) loss reserving, where aggregate methods have been the norm and approaches based on simulating (the development of) individual claims have only recently been applied to real data (\cite{APL:13, PIG:13}), simulation of individual claims has long been used in the more straightforward pricing domain; see \cite{PAP:97} for a good example. This so-called frequency-severity method is based on simulating both the number and sizes of individual claims (see also \cite{DPP:94}).

What is different in our approach compared with the traditional distribution-based simulation methods for pricing, is that we do not simply generate claim amounts from an estimated claim size distribution. Instead, we exploit the characteristics of the life/accident insurance setting by explicitly modeling the deaths occurring to policyholders and the subsequent insurance payments laid out in the contract terms; this makes it possible to accurately capture the structure of the insurance portfolio under consideration, as well as the (exact, individual) contract terms of the underlying policies, giving us the loss distribution of the primary insurer. By recording all the simulated losses individually at contract level, it is straightforward to overlay or embed the terms and conditions of the proposed reinsurance programme to arrive at the full loss distribution to the reinsurer. The contract level simulation approach is called microsimulation in this paper, in contrast to the traditional individual claims simulation methods, which make no reference to the actual contracts.

When finalizing this paper, we learned of a similar work already published. In \cite{EKH:12}, Ekheden and H\"ossjer are the first to propose a marked Poisson process type model for life catastrophes; the mechanism by which total catastrophe deaths are converted into those covered by an insurance company is also somewhat similar to the one proposed in this paper. The article \cite{EKH:12} provides an interesting analysis of a dataset larger than ours, and for different regions: our results in the first part of this paper mostly agree with their findings. However, while the authors in \cite{EKH:12} \emph{assume} that the occurrence of catastrophic events follows a homogenous Poisson process, we start by considering more general Poisson point processes, and base our choice of the specific marked Poisson process -- implying a homogenous Poisson process for the occurrence times -- on statistical tests and model comparisons. This paper therefore provides justification for the model a priori assumed and used in \cite{EKH:12}. We further extend the catastrophe model by modeling small and big events with separate mark distributions, making it possible to model the whole range of accident sizes and leading to a better description of the relatively smaller accidents.

Our paper further differs from \cite{EKH:12} in that we suggest a microsimulation approach to pricing life (re)insurance at individual contract level to capture the exact features of the underlying insurance portfolio: we don't only simulate the number of lives lost, but the actual customers that die, and the resulting contractual payments; whereas the claim distribution model used in \cite{EKH:12} is rather simplistic (gamma distribution), without reference to actual contractual amounts or contract features. The authors in \cite{EKH:12} concentrate on pricing a specific type of catastrophe Excess-of-Loss reinsurance contract on a term insurance portfolio, whereas our approach is fully general, covering all types of reinsurance contracts -- and, in principle, extending to all types of underlying life insurance products.

The outline of the paper is as follows. Using extreme value theory and statistical techniques based on it, introduced in Section~\ref{bg}, we build a model for the occurrence and sizes of catastrophic accidental deaths in Finland, using data on Finnish and Swedish populations as the basis of our analysis in Section~\ref{accdeaths}. We also discuss catastrophe risk measurement by means of the model. A microsimulation framework for reinsurance pricing is then introduced in Section~\ref{reins_pricing}, and the developed accidental death model is extended for pricing excess reinsurance. Finally, Section~\ref{conclusions} ends with summary and conclusions.

\section{Background extreme value theory} \label{bg}
In this section we will introduce the notation and concepts from extreme value theory needed for our purposes. Let $X_1, X_2, \ldots$ be a sequence of independent and identically distributed (iid) random variables with common distribution function (df) $F$, defined on a probability space $(\Omega,\mathcal{F},\P)$. The random variables (rvs for short) $X_i : (\Omega, \mathcal{F})\rightarrow (E,\mathcal{E})$ take their values in a measurable space $(E,\mathcal{E})$; in what follows, we generally take $E=\mathbb{R}$, or a subset thereof, equipped with the corresponding Borel sigma-algebra $\mathcal{B}$, as usual.

In practical applications we don't typically know the underlying distribution $F$ of the rvs $(X_i)$ whose observed realization forms our data. In using extreme value theory to model extremal events, we therefore seek an asymptotic distribution directly for the maxima, via mathematical limit arguments. These limiting distributions are then used in applications as approximations for distributions of maxima or excesses of high thresholds.

In classical extreme value theory, one is interested in the behaviour of the sample maxima, $M_n = \max(X_1,\ldots,X_n)$, $n \geq 1$. That is, one looks for normalizing sequences $c_n>0$, $d_n$ such that $M^{\ast}_{n}=(M_n-d_n)/c_n$ converges in distribution (weakly), meaning that $\P\left((M_n-d_n)/c_n\leq x\right) = F^n(c_n x + d_n) \rightarrow H(x)$ as $n\rightarrow\infty$, $x\in\mathbb{R}$, where $H$ is a non-degenerate df. An alternative equivalent condition for the convergence is\footnote{If $H(x)=0$, the limit is taken to be equal to $\infty$.}
\begin{equation} \label{eq:limgev}
n\left(1-F(c_n x+d_n)\right) \stackrel{n\rightarrow\infty}{\longrightarrow} -\ln H(x), \quad x\in\mathbb{R}.
\end{equation}
If the above hold for some non-degenerate df $H$, then the underlying distribution $F$ is said to be in the \emph{maximum domain of attraction} of $H$, denoted by $F\in\mathrm{MDA}(H)$.

One of the key results of classical extreme value theory, due to Fisher and Tippett \cite{FTI:28} and Gnedenko \cite{GNE:43}, is that if the limiting distribution $H$ exists, it must be one of three types of distributions, \emph{regardless} of the underlying distribution $F$. These three are called extreme value distributions and can be parameterized into one distribution, the \emph{generalized extreme value} (GEV) \emph{distribution}
\begin{equation} \label{eq:gev_cdf}
H_{\xi,\mu,\sigma}(x)=\exp\left\{-\left(1+\xi\frac{x-\mu}{\sigma}\right)^{-1/\xi}\right\}, \quad 1+\xi\frac{x-\mu}{\sigma}>0,
\end{equation}
with shape parameter $\xi\in \mathbb{R}$, scale $\sigma>0$ and location $\mu\in \mathbb{R}$. The case $\xi=0$ is interpreted as a limit when $\xi\rightarrow 0$. The shape parameter $\xi$ governs the tail behaviour of the distribution. For fixed $x$, it holds that $\lim_{\xi\to 0}H_{\xi,\mu,\sigma}(x)=H_{0,\mu,\sigma}(x)$ from either side, meaning that the GEV parametrization is continuous in $\xi$.

\subsection{Threshold exceedances} \label{bg_gpd}
The GEV distribution discussed above is the limiting distribution for normalized sample maxima. An alternative approach to modeling extremes is to consider all observations exceeding a certain (high) threshold. The approach is then based on the distribution of exceedances over a threshold $u$, say, which is
\begin{equation} \label{eq:excessdf}
F_u(x)=\P(X-u\leq x|X>u)=\frac{F(x+u)-F(u)}{1-F(u)}, \quad 0\leq x<x_F-u,
\end{equation}
where $X$ is a rv with df $F$, and $x_F\leq\infty$ is the right endpoint of $F$. Under the same conditions that lead to \eqref{eq:limgev}, the excess distribution $F_u$ in \eqref{eq:excessdf} can be approximated, for large $u$, by the \emph{generalized Pareto distribution} (GPD) family, originally due to Pickands (see \cite{PIC:75}):
\begin{equation} \label{eq:gpd_cdf}
G_{\xi,\beta}(x) = \left\{
	\begin{array}{rl}
		1-\left(1+\xi\displaystyle\frac{x}{\beta}\right)^{-\frac{1}{\xi}}, & \xi\neq 0, \\
		1-e^{\displaystyle-x/\beta}, & \xi=0.
	\end{array}
	\right\},
\end{equation}
where the scale parameter $\beta>0$. If the shape parameter $\xi>1/2$, the variance of GP distribution does not exist; and if $\xi>1$, the mean doesn't exist either. As in the case of GEV distribution, we again have that for fixed $x$, the GPD parametrization is continuous in $\xi$, so that $\lim_{\xi\to 0}G_{\xi,\beta}(x)=G_{0,\beta}(x)$. This is especially useful in statistical modeling.

The GPD turns out to be a natural limiting distribution for many underlying distributions $F$. In fact, it can be shown that (for some positive function $\beta(u)$) $\lim_{u\rightarrow x_F}\sup_{0\leq x<x_F-u} \left|F_u(x)-G_{\xi,\beta(u)}(x)\right|=0$ \emph{if and only if} $F\in\mathrm{MDA}(H_{\xi})$, $\xi\in\mathbb{R}$; this is the so-called Pickands-Balkema-de Haan theorem (\cite{PIC:75, BDH:74}). The result means that the distributions for which normalized maxima converge to a GEV distribution are those distributions for which the excess distribution converges to GP distribution as the threshold $u$ grows. In addition, the shape parameter $\xi$ is the same for both the limiting GEV and GP distributions.

Based on the result above, the generalized Pareto distribution can be viewed as a fundamental limiting model for the distribution of observations exceeding a high threshold. With statistical applications in mind, we get the following approximation for the excess distribution $F_u$, for large $u$: $F_u(x)=\P(X-u\leq x|X>u)\approx G_{\xi,\beta(u)}(x),$ $x>u$, where $\beta(u)=\beta+\xi u$ and $\xi$ are to be estimated from data.

Pareto distribution has long been used as a claim severity distribution in (non-life) insurance due to the good empirical fit to many insurance datasets it provides; see, e.g., \cite{PHI:85, DPP:94}. The use of generalized Pareto in the extreme value context, though a relatively standard technique by now, is a more recent development in the actuarial field. For early applications to insurance with good discussion, see \cite{ROT:97, MCN:97}.

\subsection{Point process approach} \label{bg_pp}
The discussion of the previous section only considered the magnitudes of the threshold exceedances. However, often it is natural to extend the view by explicitly including a time dimension, and to consider both the magnitudes of exceedances and the timing of them. This leads naturally to a point process view of high-level exceedances (originally introduced by Pickands \cite{PIC:71} and Smith \cite{SMI:89}) as events in space and time.

Let $(X_i)_{i=1}^{n}$ be a sequence of iid rvs with df $F$ taking values in state space $E$, and assume \eqref{eq:limgev} holds for some normalizing sequences $c_n$, $d_n$. Consider a sequence of thresholds $u_n(x)=c_n x + d_n$ for some fixed $x$ and let $Y_{i,n} = \frac{i}{n}\bm{1}_{\{X_i>u_n(x)\}}$ as in \cite{QRM:05}. The variable $Y_{i,n}$ returns either the normalized time $i/n$ of an exceedance, or zero if there's no exceedance. The \emph{point process of exceedances} $N_n(A)=\sum_{i=1}^{n}\bm{1}_{\{Y_{i,n}\in A\}}$, $n=1,2,\ldots$, with state space $E=(0,1]$, counts the exceedances of threshold $u_n$ with time of occurrence in the set $A\subset E$. Alternatively, the point process of exceedances can be written as $N_n(A)=\sum_{i=1}^{n}\bm{1}_{\left\{\left(\frac{i}{n},X_i\right)\in A\right\}}$ with two-dimensional state space $E=\left(0,1\right]\times(u_n,\infty)$.

In the context of extremes, we are again interested in the asymptotic behaviour of the point process of exceedances as $n\to\infty$. It can be shown that $N_n$ converges in distribution to a Poisson (point) process\footnote{\label{fn:ppp_def}Point process $N(\cdot)$ is a Poisson point process on $E$ with intensity measure $\Lambda$, if the following conditions hold: \hfill\\ \hphantom{aa}(i) For $A\subset E$ and $n\geq 0$, $\P\left(N(A)=n\right)=\exp\left\{-\Lambda(A)\right\}\frac{{\Lambda(A)}^n}{n!}$ if $\Lambda(A)<\infty$, and $\P\left(N(A)=n\right)=0$ if $\Lambda(A)=\infty$.\hfill\\ \hphantom{aa}(ii) For any $m\geq 1$, if $A_1,\ldots,A_m$ are mutually disjoint subsets of $E$ ($A_i\cap A_j=\emptyset, i\neq j$), then the random variables $N(A_1),\ldots,N(A_m)$ are independent.} $N$; the weak convergence is established under a topology that essentially excludes sets bordering on the lower boundary (see \cite{RES:87} and \cite[Ch.~5]{EKM:97}). The intensity measure of the limiting point process can be derived from \eqref{eq:limgev} and \eqref{eq:gev_cdf}, and is given by
\begin{equation} \label{eq:ppp_Lambda}
\begin{aligned}
\Lambda(A) &= \int_{A}\lambda(\bm{s})\dd\bm{s} = \int_{t_1}^{t_2} \int_{x}^{\infty} \lambda(y)\dd y\dd t \\ &= -(t_2-t_1)\ln H_{\xi,\mu,\sigma}(x)
	= (t_2-t_1)\left(1+\xi\frac{x-\mu}{\sigma}\right)^{-1/\xi},
\end{aligned}
\end{equation}
for sets of the form $A=(t_1,t_2)\times(x,\infty)\subset E$. The above means, among other things, that the numbers of exceedances in separate rectangles $A_1,A_2,\ldots$ are independent Poisson-distributed rvs with expected values $\Lambda(A_1),\Lambda(A_2),\ldots$.

The approaches of previous subsections can be seen as special cases of the point process approach, as all the results mentioned so far can be derived from the point process representation. The tail of the excess distribution associated with a given threshold $u$ is obtained as the conditional probability that $X_i>u+x$ given $X_i>u$:
\begin{equation*}
\bar F_u(x) = \frac{\Lambda\left((0,1)\times(u+x,\infty)\right)}{\Lambda\left((0,1)\times(u,\infty)\right)} 
	= \left(1+\frac{\xi x}{\sigma+\xi(u-\mu)}\right)^{-1/\xi},
\end{equation*}
where $\beta=\sigma+\xi(u-\mu)>0$. This is just the generalized Pareto distribution model of subsection \ref{bg_gpd}.

Similarly, the point process model can be seen to imply the GEV distribution model for maxima. Consider the event $\{M_n\leq x\}$ for some $x\geq u$: this is just the event that $N_n$ has no points in $A=(0,1)\times(x,\infty)$. The limiting probability of this event is then given by $\P(M_n\leq x)\to\P(N(A)=0)=\exp\left(-\Lambda(A)\right)=H_{\xi,\mu,\sigma}(x)$.

Futhermore, we note that for any $x\geq u$, the one-dimensional process of exceedances of the level $x$ implied by \eqref{eq:ppp_Lambda} is a homogenous Poisson process with rate $\tau(x):=-\ln H_{\xi,\mu,\sigma}(x)$.

The limiting model obtained suggests that the threshold exceedances in iid data can be approximated by a Poisson point process for a high threshold $u$. To summarize, the limiting point process model has the following properties: (i) exceedances occur as a homogenous Poisson process in time; (ii) magnitudes of excesses are iid and independent of exceedance times; and (iii) the distribution of excesses is generalized Pareto. This model is often called the Peaks-Over-Threshold or POT model (e.g., \cite{QRM:05}).

So far we have discussed the point process approach in the context of iid random variables. The results for iid sequences continue to hold for stationary sequences with small modifications, under certain conditions (see \cite{LLR:83}, the review article \cite{LER:88}, or \cite{EKM:97} for a textbook treatment). In the case of non-stationary sequences, the general theory is not helpful in practice; as noted by Smith in \cite{SMI:89}, the class of limiting distributions is much too broad to be of use in identifying parametric statistical models. In practice non-stationarity in data can be taken care of by appropriate statistical modeling: The most straightforward way to generalize the point process model discussed above is to allow the model parameters $(\xi,\mu,\sigma)$ to depend on time, or other explanatory variables.

\section{Modeling accidental deaths} \label{accdeaths}
In the following we will consider catastrophe risk relating to the life type of insurance business. More precisely, we will concentrate on the risk arising from catastrophic events causing large numbers of deaths. Our aim is to build a probabilistic model for the occurrence and sizes of these events using the theory of previous section. The data analysis, estimation and simulations in this paper are performed using the MATLAB computing language by MathWorks.

\subsection{Description of data} \label{data}
Our data consists of accidents that have occurred in Finnish and Swedish populations during the period 1910--2009, and have claimed more than three lives.\footnote{The data is based on a dataset provided by Mr. Tapani Tuominen (Kaleva Mutual Insurance Company) and collected from public sources; see, in particular, the Finnish Wikipedia (\url{fi.wikipedia.org}), class ``Onnettomuudet Suomessa'' and the Swedish Wikipedia (\url{sv.wikipedia.org}), category ``Olyckor'' and article ``Lista \"over katastrofer efter antalet d\"oda svenskar''.} Although our aim is to model the accidental deaths in Finland, when analyzing the Finnish data alone and fitting extreme value models to it, we found that the models didn't capture the right tail of the implied severity (death count) distribution well.\footnote{To save space, the analysis is not shown here, but proceeds along the lines presented below and in the appendices for the extended dataset.}
We therefore aim to improve the tail modeling by basing the statistical model estimation on an extended dataset, including Swedish accidental deaths as well.\footnote{The Swedish data is not complete but includes all the biggest accidents, and thus suffices for our purposes.} We regard combining Swedish experience with the Finnish one justifiable on the basis that Sweden is the country (area, population) that most closely resembles Finland, and the geographical and economic conditions in both countries are similar. 

The limit of three lives was set based on an examination of the numbers of accidental deaths, and to get as complete a dataset as possible, as the bigger accidents are much better documented. This censoring of the smallest observations from the data does not have an impact on the modeling of extreme death counts in practice.

For each accident, the date of occurrence and the number of deaths was picked. Accidental deaths clearly related to wars were removed from the data to avoid distorting the results. As the accident death counts between the countries are generally independent, we can use the combined dataset directly for estimation of the loss severity distribution. However, there are two accidents common to both original datasets: the sinking of cruise ship Estonia in 1994 (10 Finns and 552 Swedes died), and the Indian Ocean tsunami in 2004 (179 Finns and 543 Swedes died). We take the value from Swedish data for both accidents, but scale the tsunami deaths by the ratio of the populations (in year end 2011), to arrive at a death counth of 316: it is plausible that both (types and sizes of) accidents could happen in Finnish population, but the number of tsunami deaths is scaled down to reflect the different population sizes, and thus number of tourists.

\begin{figure}[htbp]
	\centering
		\includegraphics[width=0.80\textwidth]{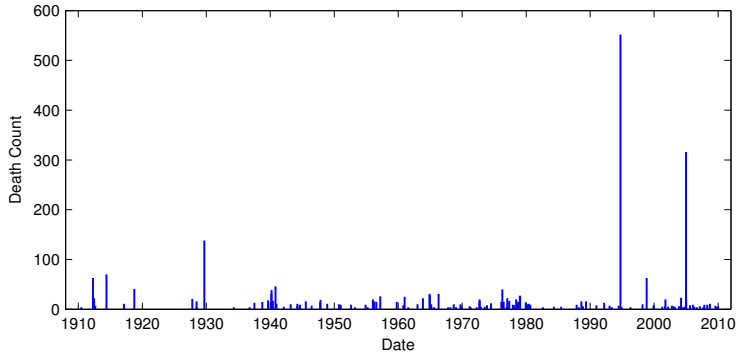}
	\caption{\small{Accidents in combined Finnish and Swedish data between 1.1.1910--31.12.2009.}} \label{fig:catplot_tot}
\end{figure}

Figure \ref{fig:catplot_tot} shows the resulting combined accidental death data, consisting of 139 accidents that claimed four lives or more, with maximum death count 552. With long series of historical observations, we have to consider whether the observations corresponding to different periods are on an equal basis. For example, when considering monetary quantities over a long period, it is essential to allow for the effects of inflation. In the context of accidental deaths, the data consists of absolute death counts, and our view is that adjustment to data is not appropriate -- possible (trend-like) features in the data should be taken care of by statistical modeling.

\subsection{Threshold selection} \label{threshold}
One of the main challenges in extreme value modeling (using either threshold exceedance or point process methods) is the selection of an appropriate threshold for fitting the model. There is inevitably a trade-off between bias and variance, as setting the threshold too low will lead to invalidity of the (limiting) GPD approximation of the excess distribution -- and hence bias in estimated parameters -- whereas setting the threshold too high will leave only few observations for the statistical model estimation, leading to high variance in parameter estimates, or even failure of the numerical estimates to converge. In practical applications, we usually try to set the threshold as low as possible, subject to the GPD providing an acceptable fit.

There are two main tools used in practice for testing whether the Generalized Pareto distribution (as implied by both the exceedance and point process methods) is a valid model for the excess distribution of observations, starting from certain threshold level $u$: the mean excess plot, and the stability of parameter estimates.\footnote{Of course, the essential diagnostic checks -- to be introduced below -- and the graphical comparison of the (tail of the) estimated distribution with the empirical distribution also tell about the goodness-of-fit for a certain threshold, \emph{once} that candidate threshold has been chosen.} Based on our analysis of the data, a threshold value of 20 is chosen; for more details, see Appendix \ref{app:threshold}.

\begin{figure}[htbp]
	\centering
		\includegraphics[width=0.40\textwidth]{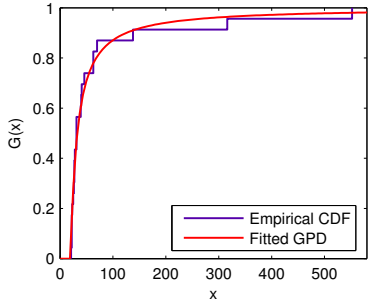}
	\caption{\small{Empirical distribution vs. fitted GP distribution.}} \label{fig:gpd_cdf_tot_u20}
\end{figure}

There are a total of 23 observations in the combined data exceeding the threshold $u=20$. Because of the small number of observations left for statistical estimation, we compared the results from using the threshold of 20 to the results obtained by using the whole dataset, corresponding to a threshold of 3: however, the value $u=3$ -- used in \cite{EKH:12} for a dataset concerning Sweden -- was found to be clearly too low for the GPD approximation to work, and is not shown here. Figure \ref{fig:gpd_cdf_tot_u20} displays the GP distribution fitted to exceedances of $u=20$. We see that the fit for the chosen threshold is quite good.

\subsection{Marked point process for accidental events} \label{mppp}
Based on our analysis, the accidental death data can be regarded iid and is well modeled by a time-homogenous Poisson process: for more details see the tests of Appendix \ref{app:homogtest} and the comparison of point process models in Appendix \ref{app:pp_models}. In this case it is convenient to interpret the two-dimensional Poisson process as a marked Poisson point process, with exceedance times the points and exceedance sizes the marks. That is, the number of points follows a (one-dimensional) Poisson process with intensity $\lambda$, and the marks are iid generalized Pareto distributed. This is the model first proposed in \cite{EKH:12}.

Given a treshold $u$, a random number $N_u$ from the sample $\bm{X}=(X_1,\ldots,X_n)$ will exceed it; re-label these observations as $\tilde{\bm{X}}=(\tilde{X}_1,\ldots,\tilde{X}_{N_u})$. The corresponding excesses are denoted by $\bm{Y}=(Y_1,\ldots,Y_{N_u})$, where $Y_j = \tilde{X}_j-u$. Since the number and mark sizes of the exceedances are assumed to be independent, the process log-likelihood can be written
\begin{equation} \label{ll_mp}
\begin{aligned}
l(\bm{\theta}; \bm{Y}) &= \ln\displaystyle\left\{\P(N_u=k)\prod_{j=1}^{k}g_{\xi,\beta}(Y_j)\right\} \\
&= k\ln\lambda - \lambda - \ln k! - k\ln\beta - \left(1+\frac{1}{\xi}\right)\sum_{j=1}^{k}\ln\left(1+\xi\frac{Y_j}{\beta}\right),
\end{aligned}
\end{equation}
where $N_u\sim\mathrm{Poi}(\lambda)$, $g_{\xi,\beta}$ is the GPD density, and $\bm{\theta}=(\lambda,\xi,\beta)$. We see that the log-likelihood \eqref{ll_mp} can be partitioned into a sum of two terms, $l(\bm{\theta}; \bm{Y})=l(\lambda;N_u)+l(\xi,\beta;Y|N_u)$, where the terms involve different parameters. This means that we can separately estimate the frequency of the exceedances (i.e., the Poisson parameter $\lambda$) and the sizes of the excesses (i.e., the GP parameters $\xi,\beta$). This marked point process representation is convenient for our purposes in section \ref{reins_pricing}, as it allows the separate simulation of event times and event sizes.

The maximum likelihood estimate for the Poisson process intensity $\lambda$ is just $\hat{\lambda}=\hat{N_u}$, the number of observed exceedances of threshold $u$ in the sample, with variance equal to the mean $\lambda$. The number of exceedances of the level $u=20$ in the combined data is 23; however, our aim is to model accidental deaths in Finnish population, not the combined experience of Finland and Sweden. To this end, we estimate the intensity from the Finnish data alone, with 15 exceedances of the threshold. To get an annualized rate, we divide the estimated intensity with the number of years of observation, $n_y=100$, to arrive at an annual intensity of $\hat{\lambda}=\hat{N_u}/n_y=0.15$.\footnote{Alternatively, we could use e.g. the average exceedance intensity of the two datasets, being $\hat{\lambda}=0.12$.} The 95 \% confidence interval is given by $[0.07, 0.28]$. 

The parameters of the mark size distribution are obtained by numerically maximizing the GP log-likelihood
\begin{equation} \label{eq:ll_gpd}
	\ln L(\xi,\beta;\bm{Y}) = \sum_{j=1}^{N_u}\ln g_{\xi,\beta}(Y_j) = -N_u\ln\beta-\left(1+\displaystyle\frac{1}{\xi}\right)\sum_{j=1}^{N_u}\ln\left(1+\xi\displaystyle\frac{Y_j}{\beta}\right),
\end{equation}
with the constraints $\beta>0$ and $1+\xi Y_j/\beta>0$ for all $j$. The maximization is done using MATLAB's {\ttfamily fminsearch} function.

The resulting parameter estimates, with approximate 95 \% confidence intervals (based on the asymptotic normality of MLEs), are given in table \ref{tbl:gpd_tot_mles}. The confidence intervals obtained from the asymptotic covariance matrix of the maximum likelihood parameter estimates are, in practice, often not very good due to the small sample size typically encountered in extreme value problems. Usually more accurate confidence intervals can be obtained by using the so-called \emph{profile likelihood}.\footnote{For each value of a parameter $\theta_i$, the profile likelihood is the log-likelihood maximized w.r.t. all other parameters. For more details, see e.g. \cite{COL:01}.} These are also shown in Table \ref{tbl:gpd_tot_mles}. The value of the shape parameter estimate $\hat{\xi}$ is over 1/2, so that the variance does not exist, and very close to 1, in which case the mean wouldn't exist either. This points to a very heavy-tailed distribution.

\begin{table}[htbp]
	\centering
	\begin{center}
	\caption{\small{GPD parameter estimates with approximate 95 \% confidence intervals.}}
	\label{tbl:gpd_tot_mles}
	\vspace{10pt}
	{\small
	\begin{tabular}{|c|c|c|c|c|}
		\hline
		\textbf{Method}  & \multicolumn{2}{|c|}{Asymptotic Std.Error} & \multicolumn{2}{|c|}{Profile likelihood} \\ \hline
		\textbf{Parameter} & MLE & 95\% CI & MLE & 95\% CI \\ \hline
		$\xi$ & 0.938 & $\left[0.189, 1.69\right]$ & 0.938 & $\left[0.457, 1.92\right]$ \\ \hline
		$\beta$ & 12.9 & $\left[6.01, 27.7\right]$ & 12.9 & $\left[6.54, 25.6\right]$ \\ \hline
  \end{tabular} }
	\end{center}
\end{table}

\begin{figure}[htbp]
	\centering
		\includegraphics[width=0.80\textwidth]{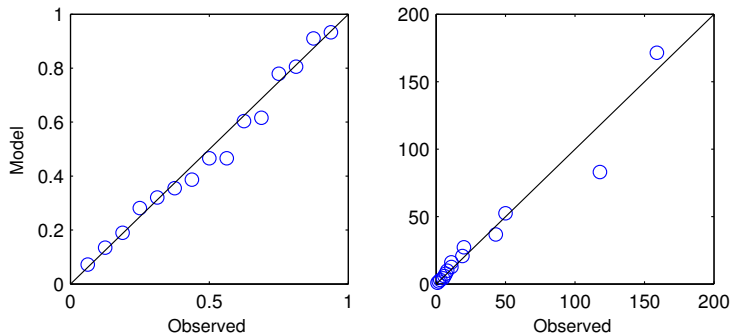}
	\caption{\small{Probability and QQ-plots; Finnish data vs. model.}} \label{fig:prob_qq_plot_totfi_u20}
\end{figure}

By plotting the empirical probabilities from the Finnish data against the ones given by the model, we get the probability plots on the left side of Figure \ref{fig:prob_qq_plot_totfi_u20}. The right side shows the corresponding quantile-quantile (QQ) plots, i.e., empirical quantiles against model quantiles. We see that the fit of the model for threshold $u=20$ is good, and without the problems encountered when the mark distribution was based on the Finnish data alone (not shown).

\subsection{Return levels and tail risk measures} \label{gpd_var}
The fitted model for accidental deaths can be used to analyze the frequency and sizes of accidents. This is conveniently done using using the concepts of return level and return period. Given the frequency of a loss event (say, a 1-in-100 year event), the magnitude of this event is called the \emph{return level}. Conversely, given the size of an event, the frequency of this kind of event is called the \emph{return period}. More precisely, let $(X_i)$ be a sequence of iid or stationary rvs with common df $F$. The return period of an event $\{X_i>u\}$ is $t_u=1/(1-F(u))$. Similarly, the return level corresponding to (return) period $t$ is given by $x_t = q_{1-1/t}(F)=F^{\leftarrow}(1-1/t)$, where $F^{\leftarrow}$ is the generalized inverse of $F$.

\begin{figure}[htbp]
	\centering
		\includegraphics[width=0.60\textwidth]{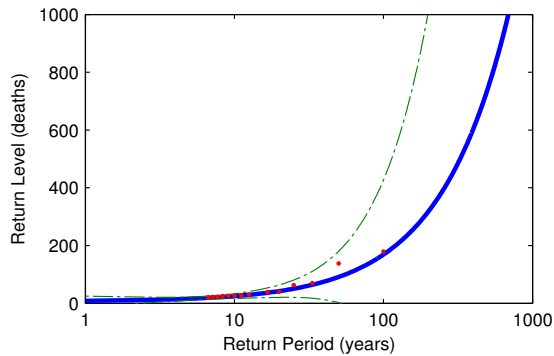}
	\caption{\small{Empirical return levels from Finnish data against the model-implied.}} \label{fig:gpd_retlev_totfi_u20}
\end{figure}

In our marked Poisson process model, the distribution of threshold excesses, conditional on an exceedance occurring, is given by the generalized Pareto distribution: $\P(X-u\leq x|X>u)=G_{\xi,\beta}(x)$, $x>0$. To calculate return levels and return periods, we also need the exceedance probability, given by $\P(X>u)=1-\exp(-\lambda_u)$, where $\lambda_u$ is the intensity of the Poisson process of exceedances related to threshold $u$. Figure \ref{fig:gpd_retlev_totfi_u20} shows the return level plot for the model, with approximate 95 \% confidence intervals.\footnote{Confidence intervals are obtained by using the delta method; see, e.g., \cite{COL:01}.} The model fit is quite good, with all the observations inside the confidence intervals.

Table \ref{tbl:gpd_retlevels_totfi} shows the model-based return levels for 10, 100, 200 and $1'000$ year accidental death events for both the delta-method based on asymptotic normality and for the profile likelihood method. The lower endpoints go negative when using the delta-method, which of course is not physically possible; there's no such problem with the profile likelihood. From the table we see that, for example, a 100-year event corresponds to 170 deaths, with 95 \% confidence interval for this point estimate ranging from 80 up to $1'100$ deaths.

\begin{table}[htbp]
	\centering
	\begin{center}
	\caption{\small{Estimated accidental death return levels with 95 \% confidence intervals.}}
	\label{tbl:gpd_retlevels_totfi}
	\vspace{10pt}
	{\small
	\begin{tabular}{|c|c|c|c|c|}
		\hline
		$\bm{u=20}$ & \multicolumn{2}{|c|}{\textbf{Delta-method}} & \multicolumn{2}{|c|}{\textbf{Profile Likelihood}} \\ \hline
		\textbf{Return Period} & \textbf{MLE} & \textbf{95\% CI} & \textbf{MLE} & \textbf{95\% CI} \\ \hline
		10 & 25 & $\left[16, 34\right]$ & 25 & $\left[22, 31\right]$ \\ \hline
		100 & 170 & $\left[-90, 430\right]$ & 170 & $\left[80, 1'100\right]$ \\ \hline
		200 & 320 & $\left[-370, 1'100\right]$ & 320 & $\left[110, 4'200\right]$ \\ \hline
		$1'000$ & $1'400$ & $\left[-4'000, 6'900\right]$ & $1'400$ & $\left[260, 1.1\cdot10^5\right]$ \\ \hline
  \end{tabular} }
	\end{center}
\end{table}

The return levels are closely connected to the popular risk measure Value-at-Risk (VaR), both being quantiles of the underlying (loss) distribution. Recall that the $t$-year return level is given by $x_t=q_{1-1/t}$; this is equal to Value-at-Risk at confidence level $\alpha$, $\mathrm{VaR}_{\alpha} = q_{\alpha}(F)$, where $\alpha=1-1/t$. This means that table \ref{tbl:gpd_retlevels_totfi} implicitly gives 1-year $\mathrm{VaR}_{\alpha}$, for $\alpha\in\{0.9, 0.99, 0.995, 0.999\}$, with associated confidence intervals. For example, the 1-year 99.5 \% VaR for accidental deaths is estimated to be 320 deaths, with upper 95 \% confidence interval for the risk measure value being $4'200$ deaths. We can also calculate Value-at-Risk explicitly by considering tail probabilities in our model: as above, with $x\geq u$,
\begin{equation} \label{eq:tailbprob}
\begin{aligned}
	\bar{F}(x) &= \P(X>x|X>u)\P(X>u) = \P(X-u>x-u|X>u)\bar{F}(u) \\
&= \bar{F}_u(x-u)\bar{F}(u) = \bar{F}(u)\left(1+\xi \displaystyle \frac{x-u}{\beta}\right)^{-1/\xi}.
\end{aligned}
\end{equation}
Here the exceedance probability $\bar{F}(u)=\P(X>u)$ is estimated as before. By inverting the above equation, we obtain
\begin{equation} \label{eq:gpdvar}
	\mathrm{VaR}_{\alpha} = F^{-1}(\alpha) = u+\frac{\beta}{\xi}\left(\left(\frac{1-\alpha}{\bar{F}(u)}\right)^{-\xi}-1\right).
\end{equation}
We can also obtain explicit formula for the other popular risk measure, the expected shortfall (ES). Essentially, expected shortfall gives the mean loss in case the loss is greater than a given high level (VaR), and is given by
\begin{equation} \label{eq:gpdes}
	\mathrm{ES}_{\alpha} = \frac{1}{1-\alpha}\int_{\alpha}^{1}q_s(F)\dd s = \frac{\mathrm{VaR}_{\alpha}}{1-\xi} + \frac{\beta-\xi u}{1-\xi},
\end{equation}
provided that $\xi<1$ (otherwise the mean of the distribution does not exist, and the integral is infinite). In the case of our estimated model the shape parameter point estimate is very close to one, $\hat{\xi}=0.938$, but expected shortfalls can nevertheless be calculated. The estimated values are tabulated in table \ref{tbl:gpd_vares_totfi}.

\begin{table}[htbp]
	\centering
	\begin{center}
	\caption{\small{Value-at-Risk and Expected Shortfall estimates for death counts.}}
	\label{tbl:gpd_vares_totfi}
	\vspace{10pt}
	{\small
	\begin{tabular}{|c|c|c|c|c|}
		\hline
		$\bm{\alpha}$ & 0.90 & 0.99 & 0.995 & 0.999 \\ \hline
		$\bm{\mathrm{VaR}_\alpha}$ & 25 & 170 & 320 & $1'400$ \\ \hline
		$\bm{\mathrm{ES}_\alpha}$ & 310 & $2'600$ & $5'000$ & $23'000$ \\ \hline
  \end{tabular} }
	\end{center}
\end{table}

We end this section with a brief remark concerning solvency regulations. In the Solvency II framework directive of the European Union, 1-year 99.5 \% Value-at-Risk is chosen as the risk measure, whereas in the Swiss Solvency Test, the risk measure used is 1-year 99 \% ES. In the context of our model, these risk measure values are estimated to be 320 and $2'600$ deaths, respectively, giving the \emph{population} level catastophe death counts.\footnote{On assessing the impact of these catastrophe events on a particular insurance company, see the next section.}

The difference between $\mathrm{VaR}_{0.995}$ and $\mathrm{ES}_{0.99}$, even though not significant for normal distribution, is remarkable here as the distribution of accidental deaths is very heavy-tailed. In this case, the capital requirement based on ES would be dramatically higher than that based on VaR, and this is true more generally for very heavy-tailed distributions. One could argue that using ES as a risk measure correctly captures the tail risk (the risk that matters most), but ultimately the risk measure (and confidence level) used in statutory solvency calculations is a policy decision, and depends on how big losses the insurers are required to be able to absorb. Although ES has theoretically better properties than VaR, the present example also shows one of the practical problems related to its use: if the mean of the distribution doesn't exist (here, if $\xi\geq 1$), the expected shortfall doesn't exist either. In the case of our model, $\xi$ is very close to one with wide confidence intervals, making the use of ES somewhat suspect.

\section{Pricing catastrophe reinsurance} \label{reins_pricing}

Having chosen a model for the occurrence of \emph{catastrophic} accidental deaths, we further extend it, in order to apply the model to reinsurance pricing in this section. The accidental death model enables explicit simulation of the times and sizes of accidents, which is key to the pricing approach presented here: combined with individual contract level data, and via a couple of intermediate steps, this allows us to obtain an accurate and consistent distribution of losses. This is due to the fact that all the features of primary insurance contracts in a reinsured portfolio, as well as the reinsurance contract terms, can be captured as they are without introducing approximations.

\subsection{Extended accidental death model} \label{ext_model}

The marked Poisson point process model of Section \ref{accdeaths} and Subsection \ref{mppp} describes the occurrence and sizes of ``big'' events (defined as those that exceed the threshold $u$). This is enough when we are interested in the extremes, i.e., the tail probabilities. However, for pricing purposes we gererally need the full distribution and therefore have to consider also the small events, for which the GPD is not a suitable model. By taking advantage of the independence assumption between event numbers and their associated mark sizes (Subsection \ref{mppp}), we can conveniently consider the frequency and severity components in turn.

Let $F_2$ denote the conditional excess distribution above threshold $u>0$, $F_2(x)=\P(X\leq x|X>u)=\P(X-u\leq x-u|X>u)=G_{\xi,\beta}(x)$, and $F_1$ the conditional distribution below the threshold, $F_1(x)=\P(X\leq x|X\leq u)$, for $x>0$. Denoting the exceedance probability of level $u$ by $p_u=\P(X>u)$, we can combine the conditional distributions into one unconditional distribution with $F(x)=(1-p_u)F_1(x)+p_u F_2(x)$. As indicated above, the excess distribution $F_2$ is GPD, while the distributional form of $F_1$ can be taken to be any count distribution, for example negative binomial, providing a good fit to data below the threshold $u$.

The ``small event'' distribution needs to be truncated from above to obtain the conditional distribution $\P(X\leq x|X\leq u)$ with support on $[0,u]$. Let $G_1$ be the full, untruncated distribution underlying $F_1$, with density (or probability function) $g(\cdot)$. In general, the density of the truncated version of $G_1$ over $[a,b]$, $-\infty<a<b<\infty$, is $f(x)=\bm{1}_{\{a\leq x\leq b\}}g(x)/(G_1(b)-G_1(a))$. Integrating (or summing) over this, we get the truncated df
\begin{equation} \label{eq:trunc_cdf}
	F_1(x) = \frac{G_1(\max(\min(x,b),a))-G_1(a)}{G_1(b)-G_1(a)}.
\end{equation}

The small and big events occur according to (homogenous) Poisson processes, $\{K_1(t),t\geq 0\}$ and $\{K_2(t),t\geq 0\}$, with intensities $\lambda_1$ and $\lambda_2$, respectively. These are mutually independent, and the Poisson intensities can be estimated based on the observed numbers of occurrences, that is, the number of events with death count within $(0,u]$ for $\lambda_1$ and $(u,\infty)$ for $\lambda_2$. Using the well known property that the sum of independent Poisson processes is again a Poisson process with rate equal to the sum of the rates of the individual processes, we get that the combined process $\{K(t),t\geq 0\}$, with $K(t)=K_1(t)+K_2(t)$, is Poisson with rate $\lambda=\lambda_1+\lambda_2$.

Conversely, let $\{K(t)\}$ be a Poisson process with rate $\lambda$. Suppose that the process is subdivided into two processes, $\{K_1(t)\}$ and $\{K_2(t)\}$: that is, arrivals of $K$ are independently sent to $K_1$ with probability $p$, and to $K_2$ with probability $1-p$. The resulting processes are each Poisson, with rates $\lambda_1=p\lambda$ and $\lambda_2=(1-p)\lambda$, and independent of each other.

The above means that we can either simulate the times of occurrence of small and big events separately; or we can simulate the occurrence of all events at once, and independently simulate the ``type'' of each event, i.e., the draw the conditional distribution from which to draw the size of the event. The results continue to hold for $n>2$ independent Poisson processes $(\{K_i(t)\}, i=1,\ldots,n)$.

\subsubsection{Combined process specification} \label{combined_proc_accdeath}
To account for the fact that our accident size data is effectively left-censored, we proceed by dividing the conditional distribution concerning the smaller death counts (denoted by $F_1$ in previous section) in two: the small death numbers on which we have data, in the interval $(u_1,u_2]:=(3,20]$, are modeled by a (truncated) negative binomial distribution; while the smallest death numbers on which we do not have data, in the interval $(u_0,u_1]:=(0,3]$, are modeled by selecting discrete probabilities for the accident sizes $\{1,2,3\}$.

The accidents exceeding threshold $u_2=20$ are modeled by the marked Poisson process of Subsection \ref{mppp}, with occurrence intensity $\lambda_3$ and mark size distribution $F_3(x)=G_{\xi,\beta}(x)$. The intensity of accidents with death count falling within interval $(u_1,u_2]$ is denoted by $\lambda_2$ and estimated from data as $\hat{\lambda}_2=0.50$, with the conditional size distribution being truncated negative binomial with parameters $r,p$ and denoted by $F_2$.\footnote{Specifically, a negative binomial distribution is first fitted into the relevant data in a left-truncated form to account for the fact that observations below and including $u_1=3$ are missing. The resulting distribution is then truncated into the interval $(u_1,u_2]=[u_1+1,u_2]$.}

Finally, we specify the intensity $\lambda_1$ for accidents with sizes in $(u_0,u_1]$ subjectively, by looking at the ratios of the frequencies of different sizes of accidents (and the individual ``intensities'' of such accidents) in the combined data, and extrapolating based on these. We end up with a combined intensity estimate of $\hat{\lambda}_1=1.63$ for the smallest accidents. The conditional probabilities $p^{1}_{i}:=\P(N=i|1\leq N\leq 3)$, $i=1,2,3$, making up the distribution $F_1$ are obtained as the ratios of individual intensities to the combined intensity $\lambda_1$. \footnote{Another way to ``force out'' the intensity $\lambda_1$ is to subjectively specify the probability of no events occurring, which is $q:=\P(N=0)=\exp\{-\lambda\}=\exp\{-(\lambda_1+\lambda_2+\lambda_3)\}$; from this we get $\lambda_1=-\ln q-(\lambda_2+\lambda_3)$ once $q$ is specified.}

We can now write the combined size distribution of accidents, given the occurrence of an accident, as $F(x)=p_1 F_1(x)+p_2 F_2(x)+p_3 F_3(x)$. Here $p_i=\lambda_i/\lambda$, with $\lambda=\sum_{i}\lambda_i$, is the probability that an event (conditional on occurring) is of type $i$, meaning that the death count associated with the event is drawn from the conditional distribution $F_i$. Table \ref{tbl:example_F_params} lists the component distributions together with their parameters. Furthermore, the unconditional distribution of event sizes can in this case be expressed as $\tilde{F}(x)=q + (1-q)F(x)$, with $q=e^{-\lambda}$ the probability of no event occurring.
\begin{table}[htbp]
	\centering
	\begin{center}
	\caption{\small{Parameters of component processes in the example.}}
	\label{tbl:example_F_params}
	\vspace{10pt}
	{\small
	\begin{tabular}{|c|c|c|c|c|}
		\hline
		\textbf{Type,} $\bm{i}$ & \textbf{Interval} & \textbf{Poisson} & \textbf{Mark (size)} & \textbf{Parameter} \\ 
		 &  & \textbf{intensity,} $\bm{\lambda_i}$ & \textbf{distribution} & \textbf{estimates} \\ \hline
		1 & $(0,3]$ & 1.63 & Discrete & $\hat{p}^{1}_{1}=0.43, \hat{p}^{1}_{2}=0.32,$ \\ 
		 & & & & $\hat{p}^{1}_{3}=0.25$ \\ \hline
		2 & $(3,20]$ & 0.50 & NegBin & $\hat{r}=1.15,$ \\ 
		 & & & (truncated) & $\hat{p}=0.182$ \\ \hline
		3 & $(20,\infty]$ & 0.15 & GPD & $\hat{\xi}=0.938, \hat{\beta}=12.9$ \\ \hline
		\emph{Total} & & \emph{2.28} & & \\ \hline
  \end{tabular} }
	\end{center}
\end{table}

\subsection{Simulation framework} \label{sim_frame}
Given the combined point process model of the previous subsection, the simulation procedure for obtaining the loss distribution for the primary insurer is outlined below. We assume a simulation period of length $T$, typically one year, and an initial number of insureds of $N_{\mathrm{ins}}$ in the insurance portfolio. For each simulation $k=1,\ldots,N_{\mathrm{sim}}$ (suppressing the dependence on $k$ to lighten notation)

\begin{enumerate}
	\item Simulate the occurrence of events within $(0,T]$ from a Poisson process with rate $\lambda$. Denote the total number of events by $K:=K(T)=K((0,T])$; if $K=0$, set total loss to zero and continue to next simulation; otherwise denote the times of events by $T_j$, $j=1,\ldots,K$.
	\item Simulate the number of deaths $N_j$ for each event $j=1,\ldots,K$. For this, first simulate the type of event $j$: with probability $p_i=\lambda_i/\lambda$, the number of deaths is drawn from distribution $F_i$, where $\lambda = \sum_{i}\lambda_i$.
	\item For each event $j$, simulate the proportion of lives lost in the event that were insured (with a policy of the type under consideration). Call this the insured percentage and denote it by $P^{I}_{j}\in [0,1]$. The number of insured deaths in event $j$ is given by $N^{I}_{j}=P^{I}_{j}N_j$.
	\item For each event $j$, simulate the proportion of insured lives lost that are covered by the insurance company. Call this the covered percentage and denote it by $P^{C}_{j}\in [0,1]$. The number of covered deaths in event $j$ is now $N^{C}_{j}=P^{C}_{j}N^{I}_{j}$.
	\item If $N^{C}_{j}>0$ and $N_{\mathrm{ins}}>0$ (i.e., not all insureds have died), randomly draw $\min(N^{C}_{j}, N_{\mathrm{ins}})$ insureds from the ones still alive. Set the state of the chosen insureds to dead, reduce $N_{\mathrm{ins}}$, terminate all the contracts still in force of each insured, and calculate the claim amounts $C_{i}^{(j)}$, $i=1,\ldots,N^{C}_{j}$ based on contract terms for each individual contract. The total claims amount from event $j$ is $C^{(j)}=\sum_i C_{i}^{(j)}$.
	\item Sum all the claims from each event $j=1,\ldots,K$ to get the total claims amount $C(T)=\sum_{j=1}^{K(T)}C^{(j)}$ for the current simulation path.
\end{enumerate}

Repeating the procedure for $N_{\mathrm{sim}}$ times, we obtain $N_{\mathrm{sim}}$ independent realizations of the primary insurer's total claims amount $C(T)$, that is, a distribution for losses. To arrive at the loss distribution for the reinsurer, we overlay or embed the rules of the reinsurance cover into the simulation steps. For instance, per occurrence aggregate limits are naturally applied to aggregate losses $C^{(j)}$ from each event, whereas per risk limits are applied to losses from each individual affected policy, $C_{i}^{(j)}$. 

The realizations of the Poisson process, or the event times, in the first step can be simulated by any of the standard techniques, such as summing exponential waiting times. A relatively efficient way is to first simulate the total number of events falling in the interval $(0,T]$, $K(T)\sim\mathrm{Poi}(\lambda T)$, and then use the well-known order statistic property of Poisson process: given the number of events $K(T)$ in $(0,T]$, the event times are distributed as uniform order statistics over $(0,T]$. That is, we may simulate iid uniform rvs $U_i\sim U(0,T]$, $i=1,\ldots,K(T)$, and order them to get a realization of the times of the Poisson process $(T_j)_{j=1}^{K(T)}$.

In step two, the accident severities are simulated from the generalized Pareto distribution, or from a truncated count distribution chosen for the smaller accidents. This can be done using the standard inverse transform method based on uniform random numbers.\footnote{Let $G$ be a distribution and $F$ a truncated version of it, truncated to interval $[a,b]$. The inverse of $F$ is given by $F^{-1}(p) = G^{-1}\left(G(a)+p\left(G(b)-G(a)\right)\right)$ for a quantile $p$. Random variables can be generated based on this.}

Moving to steps three and four, the distribution of the proportion of people dying in a particular event that are insured is assumed to be product type specific, within a given market/country. Similarly, the distribution of the proportion of insured accident victims that are insured in a given company is both product type and company specific. We propose to use the insurance coverage percentage, that is, the proportion of people in a given country/market having a specific type of insurance policy, as a proxy for the mean of the distribution of insured percentage $P^{I}$. This, or similar, figure is usually available from statistics published by government institutions or national insurance associations. In the same vein, we propose to use the market share of the insurance company (within the specific product type under consideration) as a proxy for the mean of the distribution of covered percentage $P^{C}$. The actual distributions for these proportions can be taken to be any distributions on the interval $[0,1]$ (or truncated to it) which the user regards reasonable. See also \cite{EKH:12} for another approach, which goes directly from total lives lost to customers lost without accounting for the insurance coverage within a country/market.

\subsection{Discussion on the microsimulation approach to pricing} \label{pricing_discussion}
The main advantage of the presented pricing approach, relying on individual contract level simulation, is the explicit simulation of accidents -- and ultimately the customers of an insurance company dying in these accidents -- combined with contract level data, which generally allows for more accurate distribution of losses: when simulation is done at policy level, we don't need to assume some specific distributional form for the claims, but can calculate the exact contractual amounts. Bringing the simulation to contract level also makes it possible to model features of contracts (both underlying primary, and reinsurance) which otherwise cannot be readily accounted for.

Below we list some additional advantages that the microsimulation approach offers, in addition to the general benefits (such as flexibility and the possibility to explicitly simulate order and times of events) shared with traditional frequency-severity simulation methods, and with the approach presented in \cite{EKH:12}: 

\begin{itemize}
	\item Consistent handling of random events: If an insured dies, all the policies where (s)he is the insured terminate at that point; and the contracts where (s)he is the policyholder but not the insured can either terminate or not, depending on the terms of the specific contract. Explicitly accounting for the time dimension also means that the same insured cannot die more than once (i.e., the same insured, or policy, cannot be drawn many times during a simulation path).
	\item Accounting for contract maturity: The termination of policies during the reinsurance period can be taken into account based on the maturity date of each policy.
	\item Explicit modeling and inclusion of new business: This can be very important in treaty reinsurance if the new policies incepted during the reinsurance coverage period are automatically covered by the reinsurance agreement. New policies can be included by picking policies (randomly or non-randomly) from an external data file as they become available (according to their inception date), or by simulating the number and characteristics of policies to be added.
	\item Possibility to explicitly model the sizes of total and insured populations during each simulation path in the catastrophe context: See Appendix \ref{accfor_inspop}.
	\item Modeling of additional characteristics of the accident events: For example, we can attach a certain \emph{type} to each event, and this event type can again influence other variables within the simulation, such as (the existence of) reporting delays (see subsection \ref{ibnr}) individually for each affected contract.
		\item Contract-level attribution: It is possible to directly attribute losses and reinsurance premiums down to the level of individual policy.
\end{itemize}

All the features mentioned above, except the last, contribute to producing a more accurate distribution for claims. Depending on the intended use and the features of both the reinsurance and the underlying primary insurance contracts, the specific advantages listed above can have importance. They are certainly nice from a technical point of view. Naturally, there may be cases where policy level simulation does not bring much benefits compared to a simpler individual claim, or even aggregate claim, simulation model in practice -- as always, a right tool should be chosen for each task.

The link that appears the weakest in our approach is, admittedly, the conversion of death counts from market level events into company-specific death numbers. However, we regard the adopted approach of using the insurance coverage and market share as proxies to determine the means of the affected proportion distributions as a pragmatic choice. With this approach, all the different possible outcomes are covered in principle; and if the distributions are chosen reasonably, the weights (probabilities) assigned to each outcome provide a realistic result, given a large enough number of simulations.

\subsubsection{Reporting delays and IBNR} \label{ibnr}
In contrast to P\&C (re)insurance -- where it can take a long time before a claim is reported to the insurer, and a long time thereafter before the claim is fully developed and thus the final amount known -- in insurance products offering cover for death the reporting delays are usually negligible, and don't have practical importance from modeling point of view. That is, for all practical purposes the deaths caused by accidents can usually be assumed to be reported and known without delay. There's generally no development period either, and thus no need to model the development of (new and open) claims with respect to time, as oppposed to P\&C insurance -- in fact, with these assumptions there is no such a thing as open claims.

However, there can still be cases where modeling the reporting delays and thus the IBNR (Incurred But Not Reported) claims is relevant in life/accident insurance also. The reporting delay could depend on the size of an accident, as well as on its type: for instance, only deaths related to big enough accidents might be taken to have a reporting delay, and the distribution of the delay might depend on the type of an accident. To give just one example, some of the casualties of the 2004 tsunami took some time to be identified (and some were never found).

If desired, the reporting delays can be implemented within the current simulation framework by simulating a reporting delay for each death of an insured within each accident. In the pricing context the inclusion of reporting delays does not affect the primary insurer's total claims amount for some product types (such as term insurance with specified fixed death benefits), but might affect them for others (such as savings contracts, unless the death benefit payment is tied to the savings amount at the time of death); depending on reinsurance terms, it might also affect the reinsurance recoveries; and if discounting is used or inflation is included, it has an effect, though probably not very significant in most cases. If the simulation model is used for solvency assessment of the primary insurer, instead of pricing, reporting delays mean that a claims reserve (INBR) might form at the end of the simulation period.

\subsubsection{Unified modeling of mortality risk} \label{unified_mortmodel}

The accidental death model used in this section can be easily extended to include the non-catastrophe related (``normal'') small accidents, by specifying a mortality model\footnote{This can be a stochastic mortality model, an intensity or probability function, or in the simplest case a constant accidental death intensity, for consistency estimated from such data that excludes the big accidents. As a first approximation, the pricing intensity for accidental deaths could be used.} for these, and then simulating the ``normal'' accidental deaths for each interval $(T_{i-1},T_i]$, $1\leq i\leq K(T)$, of $(0,T]$. This is done separately for each insured, and the accidental death intensity can depend on the gender, age, and other characteristics of the insured.

Furthermore, the model can be extended to include also normal -- non-accident related -- mortality in exactly the same way. This makes it possible to account for all mortality risk in a unified way within one model.

\subsection{Pricing example} \label{pricing_example}
We apply the simulation approach to rating of different reinsurance contracts, building on our case study of the Finnish accidental deaths in Section \ref{accdeaths}. We consider a portfolio consisting of $400'000$ risk life / term insurance policies, with each one belonging to different insured for illustration purposes. In the event of death, the policies pay out the risk sum (death benefit) agreed in the contract to the beneficiary. The insurance portfolio, although hypothetical, represents a realistic example of a portfolio of a Finnish life insurer. The mean risk sum in this example is around 63 kEUR, while the minimum and maximum are 5 kEUR and 10 MEUR, respectively. The total risk sum of the portfolio is approximately 25 GEUR.

\begin{figure}[htbp]
	\centering
		\includegraphics[width=0.80\textwidth]{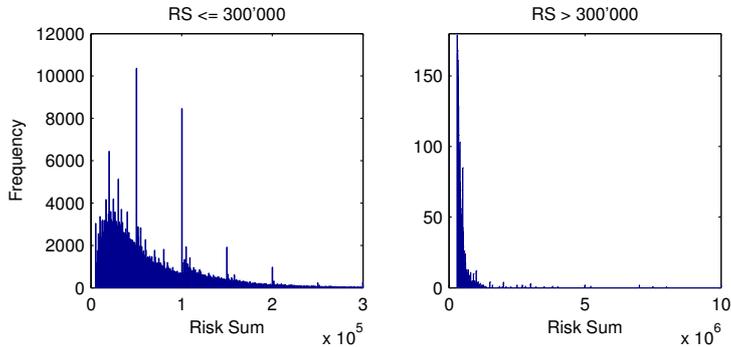}
	\caption{\small{Distribution of risk sum sizes (note the different scales).}} \label{fig:amts_hist_300k}
\end{figure}

Figure \ref{fig:amts_hist_300k} displays the distribution of risk sum sizes, separately for smaller and bigger sums to provide better impression. We observe the spikes in the number of contracts corresponding to round numbers, such as 20, 50, and 100 kEUR, among others. This is typical of real risk life portfolios, and shows that modeling the risk sums (i.e., the claim size distribution) by using a continuous distribution is not likely to yield very realistic results; at least, a number of discrete point masses would have to be mixed with a continuous distribution, but even this is likely not enough to capture the typically relatively long and sparse tail of the actual risk sum distribution.

\subsubsection{Affected proportions} \label{affected_proportions}
We use beta distribution for both the insured percentage ($P^{I}_{j}$) distribution and the covered percentage ($P^{C}_{j}$) distribution, and initially assume that these distributions remain the same for each accident $j$ during the simulation period. The mean $\mu_I$ of the $P^{I}$ distribution is taken to be the proportion of people in Finland who have a (private) risk life insurance policy, which is approximately 40 \% according to statistics. The market share of our example company is assumed to be 20 \%, and this is used as the mean $\mu_C$ of the $P^{C}$ distribution.

The mean of the beta distribution with parameters $\alpha, \beta$ is given by $\mu=\alpha/(\alpha+\beta)$; in this example we choose the parameter values $\alpha_{I}=2$ and $\alpha_{C}=0.5$, with the resulting values $\beta_{I}=3$ and $\beta_{C}=2$ as the means are fixed. The parameters for the insured distribution are chosen so that all sizes of outcomes are possible, but very low and very high values are little less likely. The covered distribution parameters, on the other hand, are chosen so that the probability of a very high value (i.e., all the insured persons covered in the company) is not too negligible, and at the same time there is a real chance that none of the insured victims are insured in the specific company.

\subsubsection{Simulation results} \label{sim_results}
To illustrate the simulation approach to pricing, we consider the following types of non-proportional reinsurance contracts\footnote{For a non-technical introduction to basics of different types of reinsurance contracts, and reinsurance in general, see \cite{SRE:13}. For a list of papers concerning (non-life) reinsurance pricing, see the reading list \cite{GWP:06}.} written on the example portfolio:

\begin{itemize}
	\item \textbf{Per Risk Excess-of-Loss (XL)}, with cedant's retention and the limit of coverage applying on a per risk (policy) basis. The reinsured layer is taken to be ``10M xs 100k'', i.e., losses in excess of the retention 100 kEUR are covered up to the limit 10 MEUR. In addition, we assume an annual aggregate limit (AAL) of 80 MEUR, that is, the total losses covered by the reinsurer are limited to this amount per annum, and an annual aggregate deductible (AAD) of 1 MEUR, meaning that the first 1 MEUR of reinsurance recoveries are retained by the ceding company.
	\item \textbf{Per Occurrence Excess-of-Loss}, with layer 50M xs 0.5M, providing a limit of coverage in excess of the retention on a per occurrence (i.e., event, accident) basis. That is, the per occurrence treaty applies to all risks affected by a single event (as defined in the reinsurance agreement\footnote{Often a minimum size is set for an event to qualify for protection, such as a minimum of 3 lives lost. Usually there's also a time limit after which consecutive losses are taken to be caused by two different events, such as 48 or 72 hours.}), providing catastrophe protection. We additionally assume that the contract includes one free reinstatement\footnote{Reinstatement means that a reinsurance cover exhausted by loss payments is replenished, i.e., brought back to full limit. Often the reinstatement takes place only after full exhaustion of the limit, and only for the remaining contract period; this is what we assume here.}, and an annual aggregate limit of 150 MEUR.
	\item \textbf{Stop-Loss}, where the reinsurer covers all the ceding insurer's losses in excess of a given amount (or a given loss ratio percentage). This is technically similar to excess-of-loss contract (aggregate XL), but the risk is now the total claims amount from the insurance portfolio / business. We consider a stop-loss contract with 20 MEUR retention and 500 MEUR limit.
\end{itemize}

\begin{figure}[htbp]
	\centering
		\includegraphics[width=0.70\textwidth]{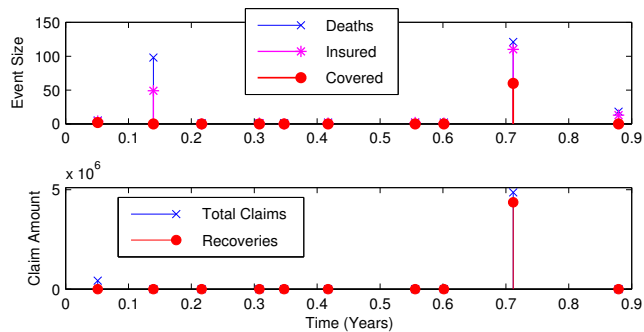}
	\caption{\small{One simulation path; number of deaths and claim amounts.}} \label{fig:sim_realization_perocc_wlegend}
\end{figure}

Without loss of generality, we consider the typical 1-year contract period ($T=1$), and run $10^6$ simulations for each reinsurance contract. Figure \ref{fig:sim_realization_perocc_wlegend} illustrates the simulation output by showing one example realization for the per occurrence reinsurance cover.

Table \ref{tbl:sim_results} shows the results from one set of simulations.\footnote{For illustration, the simulated numbers are shown without rounding, though of course these particular figures are not exact.} The means of the loss distributions are of interest in pricing, as they give expected loss costs from the corresponding contracts. Very high quantiles are also displayed to show the working of the reinsurance contracts at the extreme tail of the loss distribution.

As shown by the second column of the table, the loss cost is estimated to be approximately $6'000$ EUR for the Per Risk (Policy) contract, $46'000$ EUR for the Per Occurrence contract, and $25'000$ EUR for the Stop-Loss contract, with the specific features described above; put another way, the loss cost compared to the layer limit (sometimes the rate on line) is 0.060 \% (of the 10 MEUR per policy limit), 0.090 \% (of the 50 MEUR per occurrence limit), and 0.0050 \% (of the 500 MEUR stop-loss limit), respectively, for these contracts. Of course, these are not the reinsurance contract prices yet, as the very large uncertainty has to be taken into account in the form of a suitable risk load, based on the standard deviation perhaps, or, more appropriately, on the higher quantiles of the loss distribution; see the end of this subsection for a very brief note on the actual pricing. 

The statistics for events and event sizes, including the resulting ground-up losses on the primary insurance portfolio before reinsurance (all of which are independent of the modeled reinsurance contracts) can be used to gauge the reasonability and accuracy of individual simulation runs. In this respect, the used amount of $10^6$ simulations appears to be enough to provide consistent results for pricing purposes. However, due to the extreme heavy-tailedness of the GPD severity component, the losses in the extreme right tail can differ from simulation to simulation, and thus the results are not exactly the same each time. Note, in particular, that the lower quantiles are similar between the three simulations, but there are differences in the highest quantiles. Looking at the maximum claim amounts, we observe that in the ``per occurrence'' simulation an event happened that wiped the whole insurance portfolio, whereas the other two do not include an event of this magnitude --- however, based on other simulations not shown here, this only materially affects the utmost right tail of the loss distribution. For reinsurance pricing purposes, the maximum possible loss is usually limited, effectively mitigating the problems with estimating the very highest quantiles of the underlying (primary insurer's) loss distribution.

For a comparison, we also ran another set of simulations where the insured and covered proportions (modeled with beta distributions above) were fixed at their expected values instead, to assess the effect this random component has on the results. It turned out that the variation in the outcomes reduced, as expected, in that the values of lower quantiles of the loss distributions increased while those of higher quantiles decreased. In sum, the the loss costs for the per policy and per occurrence reinsurance contracts decreased, to around $5'000$ EUR and $41'000$ EUR, respectively, while loss cost of the stop-loss contract (with higher variation anyway) essentially remained the same.

Finally, note that we concentrated here on the estimation of the pure or net premium, corresponding to the loss cost to the reinsurer without additional loadings or margins. The gross reinsurance premium usually contains several possible elements such as expense loading, brokerage fee, ceding commission, and risk load or targer return rate. Ultimately, the actual premium rate quoted may be partly determined by other considerations, such as strategic or market conditions, or overall customer relationship. In spite of this, knowing the ``correct'' technical premium rate remains essential.

\section{Conclusions}\label{conclusions}

In the first part of this paper, we examined Poisson point process models for the occurrence and sizes of extreme accidental, or catastrophic, deaths. We provided a case study concerning accidents in Finnish population; due to the scarcity of data, we based the estimation of the (conditional) loss size distribution on combined Finnish and Swedish accident data. This approach turned out to yield good results in practice.

Based on statistical tests and model comparisons, a marked Poisson process model with independent Generalized Pareto distributed marks was chosen as the final model. This is the model originally proposed in \cite{EKH:12}; our paper provides justification for the choice. The model describes the observed big accident events well, while being simple in structure. It also allows extrapolation outside the range of data and the assessment of the risk of events more extreme than those observed so far, a key requirement for a useful catastrophe risk model. As the implied GP distribution does not describe the smaller death counts well, we adopt a pragmatic approach and model smaller accidents separately with a suitable count distribution (or several of them). The resulting combined MPP model is able to capture the whole range of accidental death counts flexibly.

With the great uncertainty related to extremes, the limitations of any statistical model have to be acknowledged. Still the fact remains that tail risk needs to be assessed, and models are needed for that purpose; the models that are ultimately used should have a sound basis. We see the catastrophe risk modeling approach presented in this paper as a tool to support decision making and to facilitate an honest assessment of the risk and uncertainty related to extreme events.

In the second part of the paper we presented a framework for reinsurance pricing, based on individual contract level simulation and using our extended accidental death model. The microsimulation approach enables the use of all the information a (re)insurer has, with the potential to provide more accurate results than traditional pricing techniques: By explicitly simulating the occurrence of accidents and the resulting deaths of insureds, the terms of each affected contract can be individually accounted for, leading to an accurate and consistent estimate of ultimate claims and outstanding liabilities. As all the claims are recorded at contract level per each accident, it is straightforward to overlay or embed the features of reinsurance contracts into the simulation. The approach makes it possible to model even complex reinsurance features without approximations. As an application, the simulation model was used to price non-propotional excess reinsurance contracts with additional aggregate features. The method presented here is tailored for life and accident catastrophe reinsurance, but could be useful in non-life contexts also.

Some of the features we discussed might be seen as unnecessary complications from a purely practical point of view, taking into account the fundamental uncertainty that is anyway related to modeling and pricing catastrophes. However, it is our view that we should try to minimize the uncertainty as much as possible where this can be plausibly done by proper (i.e., more realistic) modeling. The contract level simulation approach may also reveal risks that remain hidden when using aggregate methods, or even when simulating individual claims without direct reference to underlying contracts. Our approach does not constrain the modeler or force the use of approximations -- instead, (s)he is free to include the level of detail fit for purpose.

In addition to pricing and reinsurance design, the presented microsimulation approach is also suitable for assessing the risks of an insurance portfolio from the viewpoint of a primary insurer. This can include assessment of the need for reinsurance, as well as examining the (life) catastrophe risk from a solvency perspective.

\section*{Acknowledgements} \label{acknowledgements}
The author would like to thank Dr. Lasse Koskinen from Model IT Ltd and Aalto University for helpful comments on the paper; and Mr. Tapani Tuominen, formerly with Kaleva Mutual Insurance Company, for providing what formed the basis of the accidental death data used in this study.

\begin{landscape}

\begin{table}[htbp]
	\centering
	\begin{center}
	\caption{\small{Simulation results for the reinsurance contracts rated; total claims, reinsurance recoveries, and claims after reinsurance.}}
	\label{tbl:sim_results}
	\vspace{10pt}
	\resizebox{1.7\textwidth}{!}{
	\begin{tabular}{|c|c|c|c|c|c|c|c|c|c|c|c|c|c|c|}
		\hline
		\multicolumn{15}{|l|}{\textbf{Per Risk Excess-of-Loss}} \\ \hline
		\textbf{Quantile} & \textbf{mean} & \textbf{min} & \textbf{max} & $\bm{0.5}$ & $\bm{0.75}$ & $\bm{0.9}$ & $\bm{0.95}$ & $\bm{0.99}$ & $\bm{0.995}$ & $\bm{0.999}$ & $\bm{0.9995}$ & $\bm{0.9999}$ & $\bm{0.99995}$ & $\bm{0.99999}$ \\ \hline
		\textbf{Total Claims} & 122'197 & 0 & 4'874'273'500 & 0 & 37'200 & 161'400 & 316'100 & 1'126'100 & 1'950'650 & 8'056'450 & 15'162'050 & 65'365'200 & 142'90'1700 & 669'485'650
 \\ \hline
		\textbf{Reinsured} & 5'968 & 0 & 80'000'000 & 0 & 0 & 0 & 0 & 0 & 0 & 612'500 & 1'953'000 & 11'325'500 & 25'081'000 & 80'000'000
 \\ \hline
		\textbf{After} & 116'229 & 0 & 4'794'273'500 & 0 & 37'200 & 161'400 & 316'100 & 1'124'150 & 1'926'500 & 7'542'050 & 13'366'900 & 54'927'350 & 117'099'100 & 589'485'650
 \\ \hline
	\\
	\hline
 	\multicolumn{15}{|l|}{\textbf{Per Occurrence Excess-of-Loss}} \\ \hline
		\textbf{Quantile} & \textbf{mean} & \textbf{min} & \textbf{max} & $\bm{0.5}$ & $\bm{0.75}$ & $\bm{0.9}$ & $\bm{0.95}$ & $\bm{0.99}$ & $\bm{0.995}$ & $\bm{0.999}$ & $\bm{0.9995}$ & $\bm{0.9999}$ & $\bm{0.99995}$ & $\bm{0.99999}$ \\ \hline
		\textbf{Total Claims} & 140'590 & 0 & 25'308'264'500 & 0 & 41'600 & 172'000 & 330'000 & 1'136'800 & 1'993'900 & 8'554'150 & 15'920'850 & 77'712'450 & 128'326'600 & 639'882'200
 \\ \hline
		\textbf{Reinsured} & 45'532 & 0 & 100'000'000 & 0 & 0 & 0 & 0 & 562'700 & 1'399'350 & 7'784'200 & 14'657'350 & 62'297'750 & 100'000'000 & 100'000'000
 \\ \hline
		\textbf{After} & 95'058 & 0 & 25'208'264'500 & 0 & 41'600 & 172'000 & 330'000 & 523'700 & 602'600 & 960'800 & 1'073'600 & 10'485'000 & 49'706'400 & 539'882'200
 \\ \hline
\\
	\hline
 	\multicolumn{15}{|l|}{\textbf{Stop-Loss}} \\ \hline
		\textbf{Quantile} & \textbf{mean} & \textbf{min} & \textbf{max} & $\bm{0.5}$ & $\bm{0.75}$ & $\bm{0.9}$ & $\bm{0.95}$ & $\bm{0.99}$ & $\bm{0.995}$ & $\bm{0.999}$ & $\bm{0.9995}$ & $\bm{0.9999}$ & $\bm{0.99995}$ & $\bm{0.99999}$ \\ \hline
		\textbf{Total Claims} & 131'351 & 0 & 6'945'650'200 & 0 & 41'400 & 171'200 & 331'000 & 1'158'600 & 2'023'750 & 8'015'200 & 15'495'300 & 78'456'800 & 154'211'650 & 745'940'400
 \\ \hline
		\textbf{Reinsured} & 24'778 & 0 & 500'000'000 & 0 & 0 & 0 & 0 & 0 & 0 & 0 & 0 & 58'456'800 & 134'211'650 & 500'000'000
 \\ \hline
		\textbf{After} & 106'573 & 0 & 6'445'650'200 & 0 & 41'400 & 171'200 & 331'000 & 1'158'600 & 2'023'750 & 8'015'200 & 15'495'300 & 20'000'000 & 20'000'000 & 245'940'400
 \\ \hline
  \end{tabular} }
	\end{center}
\end{table}

\end{landscape}

\appendix 

\section{Accounting for limited insured population size} \label{accfor_inspop}
In some cases we might want to explicitly take into account the population (market) size and the size of insured population. The following example makes this clear: In the setting of our example, suppose that the total population in the insurance market considered is initially $N^{\mathrm{pop}}_{0}=5$ million people, and suppose further that $\mu^{I}_{0}=40$ \% of the population have an insurance policy of the kind we are interested in, meaning that a total of $N^{\mathrm{ip}}_{0}=2$ million people are insured for our purposes. Now if an accident happens where the death count $N_j$ is greater than $N^\mathrm{pop}_{j}-N^\mathrm{ip}_{j}=3$ million (say it's 4 million), we know that the number of insured people dying in the accident is at least $N^\mathrm{ip}_{j}-\max\{N^\mathrm{pop}_{j}-N_j,0\}$ (say, 1 million). Here $N^\mathrm{pop}_{j}$ and $N^\mathrm{ip}_{j}$ are the total and insured population sizes, respectively, just before accident $j$, at time $T_{j}-$ (with $T_{0}-=T_{0}=0$). 

Let $P^{I}_j$ be the (simulated) insured proportion in accident $j$ in one simulation, as before. In the context of the simulation framework, accounting for the limited insured population size means that we replace the number of insureds affected, $N^{I}_j=P^{I}_j N_j$, with
\begin{equation*}
	N^{I}_j=\min\{N^\mathrm{ip}_{j}, \max\{P^{I}_j N_j, N^{\mathrm{ip}}_{j}-\max\{N^{\mathrm{pop}}_{j}-N_j,0\}\}\}.
\end{equation*}
After each accident, the total and insured population sizes are reduced by the numbers of casualties, $N_j$ and $N^{I}_j$, respectively, subject to a floor of zero. As the population sizes reduce due to accidents, also the insurance coverage in the population changes (depending on who dies and who doesn't), and is given by the new proportion of insureds to total population as $\mu^{I}_{j}=\tilde{N}^{\mathrm{ip}}_{j}/\tilde{N}^{\mathrm{pop}}_{j}$ applying between $(T_j,T_{j+1}]$ (i.e., between accidents $j$ and $j+1$); here the notation means $\tilde{N}^{\mathrm{ip}}_{j}=N^{\mathrm{ip}}_{j}-N^{I}_j$, that is, the number of insured population immediately after accident $j$ (at time $T_{j}+$) --- in general, this does not need to be same as $N^{\mathrm{ip}}_{j+1}$ (at time $T_{j+1}-$) if, for instance, other deaths than those caused by (bigger) accidents are modeled, or if new business is included in the simulation model; see Section \ref{pricing_discussion}.

A procedure similar to the one described above applies also to the number of insured people, dying in a given accident, that are covered by a specific insurance company. Say that the market share\footnote{To be fully consistent, market share here should ideally be measured by the number of insureds rather than by monetary amounts, such as risk sums or savings.} of the particular company is $\mu^{C}_{0}=20$ \%, in which case the number of people (who have the insurance protection in question) covered by the company is taken to be $N^{\mathrm{cov}}_{0}=0.4$ million at the beginning. Again, if the number of insured people dying in an accident is greater than $N^\mathrm{ip}_{j}-N^\mathrm{cov}_{j}$ (1.6 million initially in our example), we know that at least a certain amount of them inevitable are insured in the company in question. Denoting by $P^{C}_j$ the random proportion of insured accident victims who are covered by the company, we thus replace the number of covered deaths, $N^{C}_j=P^{C}_j N^{I}_{j}$, with
\begin{equation*}
	N^{C}_j=\min\{N^{\mathrm{cov}}_{j}, \max\{P^{C}_j N^{I}_{j}, N^{\mathrm{cov}}_{j}-\max\{N^\mathrm{ip}_{j}-N^{I}_j,0\}\}\}.
\end{equation*}
Again, the remaining number of people covered by the company after accident $j$ is denoted by $\tilde{N}^{\mathrm{cov}}_{j}=N^{\mathrm{cov}}_{j}-N^{C}_j$, and the new market share (used as proxy for the mean of the covered percentage distribution for the next accident) is given by $\mu^{C}_{j}=\tilde{N}^{\mathrm{cov}}_{j}/\tilde{N}^{\mathrm{ip}}_{j}$ if $\tilde{N}^{\mathrm{ip}}_{j}>0$, and $\mu^{C}_{j}=0$ otherwise.

The simulations of Subsection \ref{pricing_example} were repeated with the described explicit accounting for population sizes: In our case, the obtained results did not materially differ from the previous ones without it (shown in table \ref{tbl:sim_results}).\footnote{In the present context we consider the extreme accidents from a purely technical point of view, with no aim whatsoever to account for other consequences that these kinds of catastrophes inevitable would have on the insurance company.} However, for populations that are smaller still, the explicit accounting for population sizes could be important. In these cases smaller accidents can be enough to affect the outcomes.

\section{Model selection} \label{app:model_selection}

\subsection{Tests for threshold selection} \label{app:threshold}
\begin{figure}[htbp]
	\centering
		\includegraphics[width=0.70\textwidth]{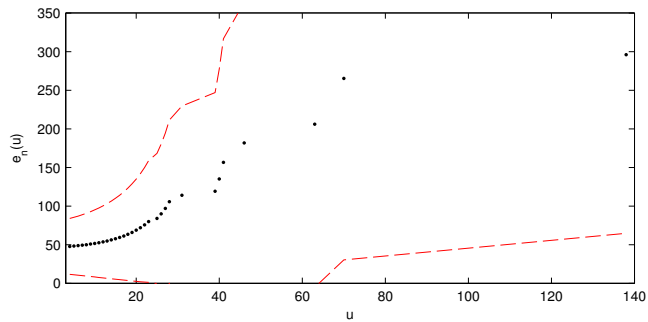}
	\caption{\small{Sample mean excess plot for the combined data.}} \label{fig:meplot_tot}
\end{figure}

\textbf{Mean excess plot.} The mean excess function $e(u)=\E(X-u|X>u)$ of a rv $X$ (if exists) gives the mean of the excess amount over $u$, conditional on $X$ exceeding $u$. For GP distribution, this is given by $e(u)=\beta(u)/(1-\xi)=(\beta+\xi u)/(1-\xi)$, and is thus a linear function of $u$; the linearity remains when considering higher thresholds $v>u$. The choice of threshold can be based on this property. In practice we use the estimate of $e(u)$ based on the sample $X_1,\ldots,X_n$, the sample mean excess function
\begin{equation}
e_n(u)=\frac{\sum_{i=1}^{n}(X_i-u)\bm{1}_{\{X_i>u\}}}{\sum_{i=1}^{n}\bm{1}_{\{X_i>u\}}},
\end{equation}
and look visually for a value $u=u_0$ from which the \emph{mean excess plot} $(u,e_n(u))$ becomes roughly linear (after allowing for sample variance); for more details, see \cite{EKM:97}. Figure \ref{fig:meplot_tot} shows the sample mean excess plot for the combined Finnish and Swedish data. We see that the plot curves up until around the value 30 and seems to straighten after that, with some kind of slight ``kink'' already at the value 20. Based on the mean excess plot alone, we might choose a threshold value of 30. Similar conclusions are drawn by considering the Finnish data alone, although in that case there's more evidence for choosing the lower threshold 20 (analysis not shown).

\textbf{Stability of parameter estimates.} Generalized Pareto distribution has a well-known ``stability property'', in the sense that the excess distribution of a GP distribution is always also GP distribution. This means, in practice, that if Generalized Pareto is an appropriate model for excesses over a given threshold $u$, it is that also for all higher thresholds $v>u$. The excess df has the same shape parameter $\xi$ but a scaling that grows linearly with threshold, $\beta_u:=\beta(u-u_0)=\beta_{u_0}+\xi(u-u_0)$, where $\beta_{u_0}$ is the scale parameter corresponding the initial threshold $u_0$. The estimate of $\xi$ based on the observed sample should therefore stay relatively constant (after allowing for sample variance, in practice stable) after threshold $u_0$, if GPD is a reasonable distributional model for excesses over $u_0$. Similarly, the scale parameter $\beta_u$ should change in an approximately linear fashion, \emph{except} if $\xi=0$; the latter inconvenience can be avoided by considering an alternative parametrization $\beta^{\ast}:=\beta_u-\xi u$, which is constant with respect to $u$.

\begin{figure}[htbp]
	\centering
		\includegraphics[width=0.80\textwidth]{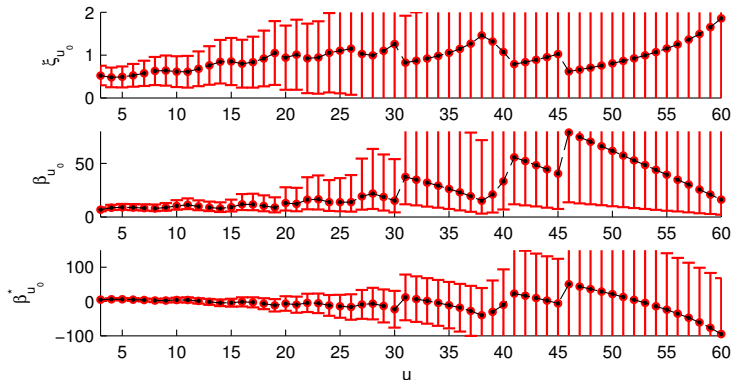}
	\caption{\small{Parameter estimates for $\xi,\beta,\beta^{\ast}$ as a function of the threshold $u$.}} \label{fig:varthr_tot}
\end{figure}

The parameter estimates for different threshold values are shown in Figure \ref{fig:varthr_tot}. In this case the threshold value 30 looks still possible, but likely not the best choice. In particular, for the pricing and risk assessment purposes we have in mind, relative stability of the model output with respect to small variations in the parameters is essential. In this light the value 30 seems too high. Instead, we take the threshold to be $u=20$. This choice is also supported if we look at the Finnish data alone.

\subsection{Testing of homogeneity} \label{app:homogtest}
Underlying both the threshold exceedance method (implicitly) and the point process approach, in its basic form, is the assumption that the exceedances of a high threshold occur according to a homogenous Poisson process, i.e., the numbers of exceedances in separate intervals are iid Poisson-distributed random variables. If this assumption is not fulfilled, the basic methods of EVT formulated for iid data are not automatically applicable, and it might be necessary to consider more general models for adequately describing the data-generating process.

\begin{figure}[htbp]
	\centering
		\includegraphics[width=0.80\textwidth]{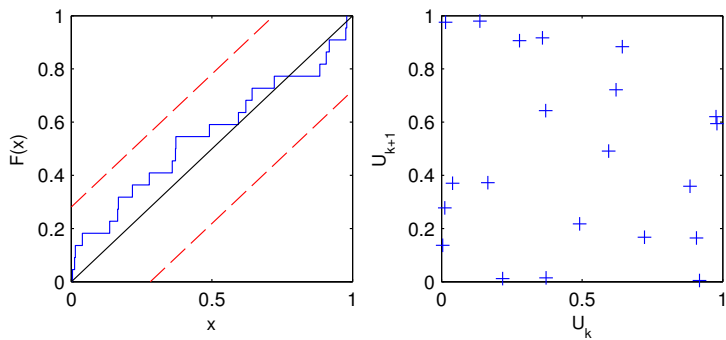}
	\caption{\small{Left: empirical vs. U(0,1) df. Right: Adjacent waiting times.}} \label{fig:iatimes_tot_cdf_uk1} 
\end{figure}

Let $\{t_i\}$ denote the exceedance times. From the basic properties of Poisson distribution we know that if the exceedance times follow a Poisson process with intensity $\lambda$, the inter-arrival (or waiting) times $T_k = t_k - t_{k-1}$ are iid exponential rvs with parameter $\lambda$. Further, the quantities $U_k = 1-\exp(-\lambda Y_k)$ are then iid and uniformly distributed on $[0,1)$. We use Kolmogorov-Smirnov (K-S) test to test the uniformity of the sample $(U_k)$: the null hypothesis that the data is uniformly distributed is not rejected. Left panel of Figure \ref{fig:iatimes_tot_cdf_uk1} shows the comparison of the empirical distribution function with the reference distribution $U(0,1)$, together with 95 \% confidence intervals. In the present case the K-S test does not have much power to reject the null hypothesis due to the small number of observations -- this is demonstrated by the wide confidence intervals. However, the figure shows no particular evidence against the uniformity of $(U_k)$ and hence the Poisson-distributedness of the exceedance times.

As in \cite{OGA:88}, the independence of inter-arrival times can also be tested by plotting the transformed waiting times $U_{k+1}$ against preceding waiting times $U_k$. If the (adjacent) inter-arrival times are independent, the points $\{(U_k,U_{k+1}) : k=1,\ldots,n\}$ should be uniformly distributed on the unit rectangle $[0,1)\times[0,1)$. Right panel of Figure \ref{fig:iatimes_tot_cdf_uk1} contains such a plot. The scarcity of points makes the interpretation somewhat difficult, but there doesn't seem to be any structure.\footnote{The conclusions of the appendix are reinforced by looking at the whole data, as the exceedance times of the threshold $u=3$ already seem to follow a homogenous Poisson process (analysis not shown).}

\subsection{Comparing point process models} \label{app:pp_models}
Based on the tests of the previous appendix, the accidental death data can be taken to be generated by a time-homogenous Poisson process. However, we will further investigate the existence of possible trends -- and whether the model can be improved by including trend components -- by fitting nonhomogenous point processes to the data.

We start by considering the (time-homogenous) POT model. This establishes the base case to which other models are compared. The points of the process $N$ consists of points $(T_j,\tilde{X}_j)\in E=(0,n]\times(u,\infty)$ from the underlying sequence of rvs $(X_1,\ldots,X_n)$ that exceed the chosen threshold $u$. The realization of the point process $N$ is thus a set of points $\{(T_j,\tilde{X}_j) : j=1,\ldots,N_u\}$, where $N_u$ is the number of exceedances and $T_j$ and $\tilde{X}_j$ represent the time and size of an exceedance, respectively.

With the intensity of the Poisson point process $N$ at a point $(t,x)\in E$ being
\begin{equation} \label{eq:ppp_lambda}
\lambda(t,x)=\frac{1}{\sigma}\left(1+\xi\frac{x-\mu}{\sigma}\right)^{-1/\xi-1},
\end{equation}
whenever $(1+\xi(x-\mu)/\sigma)>0$, and $\lambda(t,x)=0$ otherwise (see \eqref{eq:ppp_Lambda}), the likelihood based on the observed exceedance times $t_1,\ldots,t_k$ and sizes of exceedances $\tilde{\bm{x}}=(\tilde{x}_1,\ldots,\tilde{x}_k)$ (with $N_u=k$) is
\begin{equation} \label{eq:L_pot}
L(\bm{\theta};\tilde{\bm{x}}) = \exp\left\{-\Lambda\left(\left(0,n\right]\times(u,\infty)\right)\right\}\prod_{i=1}^{k}\lambda(\tilde{x}_i) \\
	= \exp\left\{-n\tau(u)\right\}\prod_{i=1}^{k}\lambda(\tilde{x}_i),
\end{equation}
where $\bm{\theta}=(\xi,\mu,\sigma)$ and $\tau(x)=-\ln H_{\bm{\theta}}(x)$ (see subsection \ref{bg_pp}). The log-likelihood is therefore given by
\begin{equation} \label{eq:ll_pot}
l(\bm{\theta};\tilde{\bm{x}}) = -n\left(1+\xi\displaystyle\frac{u-\mu}{\sigma}\right)^{-1/\xi}+\sum_{i=1}^{k}\ln\left\{\displaystyle\frac{1}{\sigma}\left(1+\xi\frac{\tilde{x}_i-\mu}{\sigma}\right)^{-1/\xi-1}\right\},
\end{equation}
when $1+\xi(\tilde{x}_i-\mu)/\sigma>0$ for all $i=1,\ldots,k$. The parameter estimates are obtained by maximizing the (log-)likelihood numerically. For more details on point process intensities and likelihood functions, see e.g. \cite[Ch.~7]{DVJ:03}.

Fitting the POT model to the accidental death data using treshold $u=20$, we get the MLEs
\begin{equation*}
\hat{\bm{\theta}}=(\hat{\xi}, \hat{\mu}, \hat{\sigma})=(0.938, 9.72, 3.25),
\end{equation*}
with the maximum log-likelihood being $-160.2$. The above implies a very heavy-tailed distribution with infinite variance. We can check that the parameters correspond to the estimates $\hat{\xi}$ and $\hat{\beta}$ of the marked point process formulation in section \ref{mppp} (using the relationship $\beta=\sigma+\xi(u-\mu)$ between the models) .

\subsubsection{Non-homogenous point processes} \label{pp_models_nonhomog}
Let $N$ be a non-homogenous Poisson point process with intensity
\begin{equation*} 
\lambda(t,x)=\frac{1}{\bm{\sigma}(t)}\left(1+\bm{\xi}(t)\frac{x-\bm{\mu}(t)}{\bm{\sigma}(t)}\right)^{-1/\bm{\xi}(t)-1}.
\end{equation*}
We consider the following parametrizations:
\begin{itemize}
	\item Model $\mathcal{M}_1$: $\quad \bm{\theta}_1=(\xi,\bm{\mu}(t),\sigma)$, where $\bm{\mu}(t)=\kappa_0+\kappa_1 t$
	\item Model $\mathcal{M}_2$: $\quad \bm{\theta}_2=(\xi,\mu,\bm{\sigma}(t))$, where $\bm{\sigma}(t)=e^{\kappa_0+\kappa_1 t}$
\end{itemize}
These are compared to the homogenous model of previous section, denoted by $\mathcal{M}_0$, to examine whether trend components can improve the model performance. Fitting the models by maximum likelihood, we get the following parameter estimates:
\begin{equation*}
\begin{aligned}
	\hat{\bm{\theta}_1} &= (\hat{\xi}, \hat{\kappa}_0, \hat{\kappa}_1, \hat{\sigma})=(0.932, 10.7, -0.023, 3.30), \\
	\hat{\bm{\theta}_2} &= (\hat{\xi}, \hat{\mu}, \hat{\kappa}_0, \hat{\kappa}_1)=(0.941, 9.77, 1.29, -0.0024).
\end{aligned}
\end{equation*}
The log-likelihood at the maximums are in both cases approximately $-160.1$. As model $\mathcal{M}_0$ is a special case of the other models (with $\kappa_1\equiv 0$), we can use the standard likelihood ratio (LR) test to compare the models. According to the test, including a trend to the location or scale parameter does not improve the fit of the model in a statistically significant way -- in fact, it barely improves it at all. Goodness-of-fit tests confirm this.


\begin{thebibliography}{99}
	
\bibitem{APL:13} K.~Antonio and R.~Plat, \emph{Micro-level stochastic loss reserving for general insurance}, Scandinavian Actuarial Journal (2013), Published online 17 May 2013.

	\bibitem{BDH:74}
	A.A. Balkema and L.~de~Haan, \emph{Residual life time at great age}, Annals of
		Probability \textbf{2} (1974), 792--804.

	\bibitem{COL:01}
	S.G. Coles, \emph{An introduction to statistical modeling of extreme values},
		Springer, New York, 2001.

	\bibitem{DVJ:03}
	D.J. Daley and D.~Vere-Jones, \emph{An introduction to the theory of point
		processes: Volume i: Elementary theory and methods}, 2nd ed., Springer, 2003.

	\bibitem{DPP:94}
	C.D. Daykin, T.~Pentik\"ainen, and M.~Pesonen, \emph{Practical risk theory for
		actuaries}, Chapman and Hall, London, 1994.

	\bibitem{EKH:12}
	E.~Ekheden and O.~H\"ossjer, \emph{Pricing catastrophe risk in life
		(re)insurance}, Scandinavian Actuarial Journal (2012), Published online 20
		Aug 2012.

	\bibitem{EKM:97}
	P.~Embrechts, C.~Klüppelberg, and T.~Mikosch, \emph{Modelling extremal events
		for insurance and finance}, Springer, Berlin, 1997.

	\bibitem{FTI:28}
	R.A. Fisher and L.H.C. Tippett, \emph{Limiting forms of the frequency
		distribution of the largest or smallest member of a sample}, Proceedings of
		the Cambridge Philosophical Society \textbf{24} (1928), 180--190, Available
		online at: \url{http://hdl.handle.net/2440/15198}.

	\bibitem{GWP:06}
	M.~Flower, \emph{A review of papers relevant to non-life reinsurance (with a
		liability focus)}, General Insurance Convention, Institute and Faculty of
		Actuaries, 2006.

	\bibitem{GNE:43}
	B.V. Gnedenko, \emph{Sur la distribution limité du terme d'une série
		aléatoire}, Annals of Mathematics \textbf{44} (1943), no.~3, 423--453.

	\bibitem{LLR:83}
	M.R. Leadbetter, G.~Lindgren, and H.~Rootzén, \emph{Extremes and related
		properties of random sequences and processes}, Springer, Berlin, 1983.

	\bibitem{LER:88}
	M.R. Leadbetter and H.~Rootzén, \emph{Extremal theory for stochastic
		processes}, Annals of Probability \textbf{16} (1988), 431--478.

	\bibitem{MCN:97}
	A.J. McNeil, \emph{Estimating the tails of loss severity distributions using
		extreme value theory}, ASTIN Bulletin \textbf{27} (1997), 117--137.

	\bibitem{QRM:05}
	A.J. McNeil, F.~Rüdiger, and P.~Embrechts, \emph{Quantitative risk management:
		Concepts, techniques and tools}, Princeton University Press, Princeton and
		Oxford, 2005.

	\bibitem{OGA:88}
	Y.~Ogata, \emph{Statistical models for earthquake occurrences and residual
		analysis for point processes}, Journal of the American Statistical
		Association \textbf{83} (1988), 9--27.

	\bibitem{PAP:97}
	D.E. Papush, \emph{A simulation approach in excess reinsurance pricing},
		Casualty Actuarial Society Forum, Casualty Actuarial Society, Spring 1997,
		pp.~1--30.

	\bibitem{PHI:85}
	S.W. Philbrick, \emph{A practical guide to the single parameter pareto
		distribution}, Proceedings of the Casualty Actuarial Society, no. LXXII,
		Casualty Actuarial Society, 1985, pp.~44--84.

	\bibitem{PIC:71}
	J.~Pickands, \emph{The two-dimensional poisson process and extremal processes},
		Journal of Applied Probability \textbf{8} (1971), 745--756.

	\bibitem{PIC:75}
	J.~Pickands, \emph{Statistical inference using extreme order statistics}, Annals of
		Statistics \textbf{3} (1975), 119--131.

	\bibitem{PIG:13}
	M.~Pigeon, K.~Antonio, and M.~Denuit, \emph{Individual loss reserving with the
		multivariate skew normal framework}, ASTIN Bulletin \textbf{43} (2013),
		399--428.

	\bibitem{SRE:13}
	Swiss Re, \emph{The essential guide to reinsurance}, Swiss Reinsurance Company,
		Available online at:
		\url{http://media.swissre.com/documents/The_essential_guide_to_reinsurance_updated_2013.pdf}.

	\bibitem{RES:87}
	S.I. Resnick, \emph{Extreme values, regular variation, and point processes},
		Springer, New York, 1987.

	\bibitem{ROT:97}
	H.~Rootzén and N.~Tajvidi, \emph{Extreme value statistics and wind storm
		losses: a case study}, Scandinavian Actuarial Journal \textbf{1} (1997),
		70--94.

	\bibitem{SMI:89}
	R.L. Smith, \emph{Extreme value analysis of environmental time series: An
		application to trend detection in ground-level ozone}, Statistical Science
		\textbf{4} (1989), 367--393.

\end{thebibliography}
\end{document}